\documentclass[preprint2]{aastex631}                      

\newcommand\vv{{\mathrm v}  }
\newcommand\vsini{\mathrm{v}\sin\mathrm{i}}
\begin{document}

\title{Transitory tidal heating events and their impact on cluster isochrones}
\author{S. Jane Arthur}
\affil{Instituto de Radioastronom\'{\i}a y Astrof\'{\i}sica, Universidad Nacional Aut\'onoma de M\'exico, Antigua Carretera a P\'atzcuaro \#8701, Ex-Hda. San Jos\'e de la Huerta, Morelia, Michoac\'an, C.P. 58089, M\'exico,
j.arthur@irya.unam.mx}

\author{Gloria Koenigsberger}
\affil{Instituto de Ciencias F\'{\i}sicas, Universidad Nacional Aut\'onoma de M\'exico,
Ave. Universidad S/N, Cuernavaca, Morelos, 62210, M\'exico, gloria@icf.unam.mx}
\affil{Department of Astronomy, Indiana University, Bloomington, IN}

\author{Kristin Brady}
\affil{Department of Astronomy, Indiana University, Bloomington, IN}

\author{Diana Estrella-Trujillo}
\affil{Instituto de Ciencias F\'{\i}sicas, Universidad Nacional Aut\'onoma de M\'exico,
Ave. Universidad S/N, Cuernavaca, Morelos, 62210, M\'exico, Universidad Nacional Aut\'onoma de M\'exico,
dtrujillo@icf.unam.mx}

\author{Catherine Pilachowski}
\affil{Department of Astronomy, Indiana University, Bloomington, IN, cpilacho@indiana.edu}

\begin{abstract}
The kinetic energy in tidal flows, when converted into heat, can affect the internal structure of a
star and shift its location on a color-magnitude diagram from that of standard models.  In this paper 
we explore the impact of injecting heat into stars with masses near the main sequence turnoff mass 
(1.26~$M_\odot$) of the open cluster M67.  The heating rate is obtained from the tidal shear
energy dissipation rate which is calculated from first principles by simultaneously solving the 
equations that describe orbital motion and the response of a star's  layers to the gravitational, 
Coriolis, centrifugal, gas pressure and viscous forces.  The stellar structure models are computed 
with MESA. We focus on the effects of injecting heat in pulses lasting 0.01~Gyr, a timeframe 
consistent with the synchonization timescale in binary systems.  We find that the location of the tidally 
perturbed stars in the M67 color-magnitude diagram is shifted to significantly higher luminosities and 
effective temperatures than predicted by the standard model isochrone and include locations corresponding 
to some of the Blue Straggler Stars. Because tidal heating takes energy from the orbit causing it to shrink,  
Blue Straggler Stars could be merger or mass-transfer progenitors as well as products of these processes.   

\end{abstract}


\section{Introduction} \label{sec:intro}

The Hertzprung-Russell Diagram (HRD) is a fundamental diagnostic tool in the study of astrophysical
problems.  Through the mediation of theoretical models, it allows information to be gleaned on the 
internal structure and evolutionary state of any star for which an effective temperature ($T_\mathrm{eff}$) 
and absolute luminosity ($L$) are known.  The theoretical models that are currently in use have
been developed over more than 7 decades and incorporate a wide variety of physical processes, 
including nuclear fusion, energy transport, plasma physics, and hydrodynamics, many of which have 
been added as their relevance becomes apparent or when there are advances in computational infrastructure. These models have, in general, been very successful in reproducing observational             
properties of stars, stellar associations and clusters.  There are, however, exceptions: stars that do not fit into the established patterns.

Cluster HRDs systematically include some stars whose $(T_\mathrm{eff},L)$ coordinates are inconsistent with
the general cluster properties. The most striking example is the Blue Straggler Stars (BSSs),
objects that are hotter and more luminous than the main-sequence turn-off (MSTO), and which should no longer be present.
These outliers point to physical processes whose relevance has been underestimated or neglected 
in the standard models.  For example, \citet{2017ApJ...845L..17C} show that the magnitude of dissipative 
heating in strongly stratified convecting fluids is not negligible, which could significantly 
alter the internal stellar structure but is not currently incorporated in model calculations.

In recent years, binary interaction effects have been invoked in order to explain stars located
at anomalous HRD locations.  The BSSs, for example, have been suggested to result from mergers
\citep{1976ApL....17...87H, 1989AJ.....98..217L} or mass exchange \citep{1964MNRAS.128..147M, 
2006ApJ...647L..53F, 2009Natur.457..288K}. However, additional binary interaction effects exist 
that can affect a star's internal structure and thus its HRD location. \citet{2016RMxAA..52..113K}  
suggested that, if tidal shear energy dissipation in an asynchronously rotating binary star is fed 
into the internal layers as heat, this would cause the star to increase its radius.  
\citet{2023A&A...670A..44E} demonstrated that if the tidal heating is injected into a stellar model,  
the reaction of the star is indeed to increase its radius but, in addition, it becomes hotter and 
more luminous than the corresponding standard model, and its evolutionary path crosses the HRD regions
occupied by BSSs. In this paper we take this experiment one step further, focusing it on the
old open cluster Messier~67. 

Shear energy dissipation becomes important when a star possesses a significant differential 
rotation structure.  Such a structure develops as a consequence of evolutionary processes when,
for example, the core contracts and the envelope expands as the star reaches the MSTO.  It is also
induced when the stellar rotation rate differs from the orbital angular velocity in a binary system. 
Asynchronously rotating systems should be found among the stars that are evolving off the main sequence
as well as systems formed by two- or three-body tidal capture, such as those prevalent at the center 
of globular clusters \citep{2009Natur.457..288K}.   

In this paper we examine the behavior of stars with masses near the MSTO of the M67 cluster ($1.26 M_\odot$).
In Section~\ref{sec:method} we describe the method used to compute the tidal shear energy dissipation 
rates and the stellar structure.  In Section~\ref{sec:results} we show the manner in which the structure of the
heated models differs from the standard models and the manner in which a star reacts when heating is
introduced and then turned off. The application of these models to the M67 cluster is described in Section~\ref{sec:M67} and recent observational results are discussed in Section~\ref{sec:disc}. Finally, we summarize and present our conclusions in Section~\ref{sec:conclusions}.
An appendix contains supporting information.
 
\section{Method} \label{sec:method}

\begin{deluxetable*}{clc}
\tablecaption{Model input parameters \label{table_input_parameters}}
\tablecolumns{3}
\tablewidth{0pt}
\tablehead{\colhead{Parameter} & \colhead{Description}  & \colhead{Value} }
\startdata
$P_\mathrm{orb}$     &Orbital period (d)                              &  1.44         \\
$e$           &Orbital eccentricity                            & 0             \\
$m_1$         &Perturbed star mass (${\rm M_\odot}$)           &1.2            \\
$m_2$         &Companion star mass (${\rm M_\odot}$)           &(a)           \\
$R_1$         &Initial radius (${\rm R_\odot}$)                &(a)           \\
$\beta_0$   & Synchronicity parameter $\omega_0/\Omega_0$       &(a)           \\
$n$           &Polytropic index                                &2.2            \\
$\Delta R/R_1$  &Layer thickness                              &0.06             \\
$N_r$         &Size of radial grid                             & 10           \\
$N^\mathrm{eq}_\varphi$&Size of longitude grid at the equator         &200           \\
$N_\theta$    &Size of latitude grid (one hemisphere)          &20             \\
$N_\mathrm{cycle}$      &Duration of the run (orbital cycles)            & 50$^{(b)}$   \\
$N_\mathrm{tt}$      &Number of orbital phases within each cycle      & 40$^{(c)}$   \\
$\mbox{Tol}$         & Tolerance for the Runge-Kutta integration      &10$^{-7}$     \\
\enddata
\tablenotetext{}{(a) See Table~\ref{tab:cases_run}.  (b) Some systems display super-orbital
periodicities which makes it important to specify the particular orbital cycle being analyzed.
(c) Asynchronous binaries display tidally induced oscillations.}
\end{deluxetable*}

\subsection{Tidal shear energy dissipation rate} \label{sec:TIDES}

We perform a calculation from first principles \citep{Moreno:2011jq, 2021A&A...653A.127K,  
2023A&A...670A..44E}.  It consists of simultaneously solving the orbital motion and the equations 
of motion of a 3D grid of volume elements that cover the central region of the tidally perturbed star, 
which we refer to as the {\it primary}.  This central region, referred to as the core, rotates as 
a solid body with constant angular velocity $\omega_0$ in an inertial reference frame, and its rotation 
axis is perpendicular to the orbital plane.  It does not coincide with the central nuclear burning 
region, and in all our models it includes a significant portion of the radiative envelope.  The equations 
of motion  are solved in the reference frame with origin in the center of
the primary and rotating at the rate of the companion's orbital motion, and include gravitational, 
centrifugal,  Coriolis, gas pressure gradient and viscous forces. A seventh order Runge-Kutta integrator 
is used to solve the set of equations. The companion is considered to be a point-mass source and its 
orbital plane is coplanar with the primary star's equator.  The code, in its latest version, is named 
TIDES-nvv and is publicly available.\footnote{Moreno, E., \& Koenigsberger, G.  2024, Tidal Interactions 
with Dissipation of Energy due to Shear version nvv, Zenodo. https://doi.org/10.5281/zenodo.10799022}

The rate of energy dissipation per unit volume, $\dot{E}_\mathrm{V}$ depends on the mass density $\rho$, 
viscosity $\nu$ as well as the velocity gradients in the radial, latitudinal and longitudinal directions across 
neighboring volume elements \citep{McQuarrie1976}.   We use the formulation given in  
\citet{Moreno:2011jq}, in which only the gradients in angular velocity are implemented.
Also, because the azimuthal perturbations are an order of magnitude larger than those in the
radial and polar direction \citep{1981ApJ...246..292S,2009ApJ...704..813H}, the calculations performed
in this paper only consider these perturbations.



The viscosity is non-isotropic and time-dependent. It is assumed to be dominated by a turbulent term 
whose value is computed for each volume element interface following the formulation of 
\citet{1987flme.book.....L}: 

\begin{equation}
 \nu_\mathrm{turb} =\lambda \ell_\mathrm{t} \Delta u_\mathrm{t},         \label{eq_Landau_Lifshitz}
\end{equation}

\noindent where $\ell_\mathrm{t}$ is the characteristic length of the largest eddies that are associated
with the turbulence,  $\Delta u_\mathrm{t}$ is the typical average velocity variation of the flow over
the length, $\ell_\mathrm{t}$.  $\lambda$ is a proportionality parameter that lies between zero
and one and which can be interpreted as the fraction of kinetic energy that is converted into heat.

The TIDES calculation reproduces the oscillatory properties of the tidal flows and the evolving differential
rotation structure that is established as angular momentum is transported between stellar layers in response
to the tidal torque. The time-marching TIDES algorithm is valid for binary stars with arbitrary rotation 
velocity and eccentricity, as long as neighboring grid elements retain contact over at least $\sim$80\% of 
their surface and the centers of mass of two adjoining grid elements do not overlap.  However, it neglects 
buoyancy effects, heat and radiation transfer, fluid dynamics microphysics, diffusion and
advection.  Also,  because the energy dissipation rates are not fed back into the system, the model
does not compute the modified internal structure that would result from this feedback. Hence, it is
valid only for studying the short-term (i.e., orbital timescales) behavior under the influence of tides.

 The input parameters are described in Table~\ref{table_input_parameters}.  The orbital period,
eccentricity, primary and companion masses are labelled $P$ and $e$, $m_1$ and $m_2$, respectively.
The primary's structure is assumed to be polytropic with an index $n$.  Its initial unperturbed radius 
is $R_1$, and its initial rotation structure is that of uniform rotation with an angular velocity $\omega_0$ 
in the inertial reference frame.  We define the parameter $\beta_0=\omega_0/\Omega_0$ which establishes 
the initial rotation state.  Here, $\Omega_0$ is the orbital angular velocity at a reference point in 
the orbit. In circular orbits, $\beta_0$ is constant but for $e\neq0$ it varies over the orbital cycle, in
which  case, $\Omega_0$ is chosen to be the orbital angular velocity at the time of periastron. 
In general,  the layers above the core respond to the forces in the system and over time develop a 
differential rotation structure.  The oscillatory tidal field is superposed on the differential rotation 
structure (see, for example, Fig.~5 in \citealp{2021A&A...653A.127K}).  The only case in which such a 
differential rotation structure does not develop is when $e=0$ and $\beta_0=1$, that is, in circular orbits 
in which the stellar rotation rate equals the orbital rate.

The computational input parameters are the grid size ($N_r \times N_\varphi \times N_\theta$),
the thickness of each layer ($\Delta R/R_1$), the number of orbital cycles over which
the computation is run ($N_\mathrm{cycle}$) and the tolerance for the Runge-Kutta integration.
The triad $(r, \varphi, \theta)$ correspond, respectively, to the distance from the stellar
center,  the azimuth angle and the colatitude. The angle $\varphi=0$ is measured from the line
joining the centers of the two stars in the direction of the orbit.\footnote{Note that because the
solution of the equations of motion is performed in the reference frame rotating with the binary
orbit ($S^\prime$ in the notation of Moreno et al. 2011), these variables correspond to the primed variables
but for the purposes of this paper, this notation is not required and thus omitted.} 

For the nominal calculations presented in this paper, we chose the primary mass $m_1=1.2\,M_\odot$, which 
is slightly smaller than the M67 turnoff mass,  $m_\mathrm{MSTO}=1.26\,M_\odot$.  The orbital period is 1.44~d, the same as 
in \citet{2023A&A...670A..44E}. We probed stars having radii $R_1/R_\odot=0.97$, 1.2, 1.32 and 1.50,
corresponding to different evolutionary times. The primary's rotation velocity was assumed to always be 
subsynchronous with $\beta_0=0.7$, which corresponds to $\vv_\mathrm{rot}/\mbox{(km\,s$^{-1}$)}=24$, 29, 32, 36 for
the above listed radii, respectively.  These values are significantly larger than typical projected 
rotation speeds in M67 which lie in the range $\vv_\mathrm{rot} \sin \mathrm{i} \sim$5--8\,km\,s$^{-1}$ \citep{2001A&A...375..851M}
but these cited values likely correspond to mostly single stars. Short-period binaries are expected to 
rotate much faster because synchronization times ($\sim10^7$~yr) are orders of magnitude smaller than 
the main-sequence lifetime.  Furthermore, \citet{2020AJ....160..169N} find 113 stars with $\vsini > 120$~km\,s$^{-1}$ 
in NGC~7789. We note in passing that these very rapid rotators are hot and blue.

Energy dissipation rates were computed for  $m_2/M_\odot=0.8$ and 0.4 companions while holding the
other parameters constant.  These models are labelled Case~1 and Case~2, respectively in Table~\ref{tab:cases_run}.
Each case is subdivided into cases b-d, corresponding to the different adopted values of $R_1$, which are
listed in Column~3 of this table.  We also computed a set of models with $m_2=0.4\,M_\odot$ and $\beta_0=1.1$, which
is a super-synchronously rotating model.  This is Case~3.


The energy dissipation rate per unit volume $\dot{E}_V$ is a time-dependent quantity due to the tidal
oscillations and, in eccentric orbits, due to the varying conditions between periastron and apastron.  Thus, 
$\dot{E}_\mathrm{V}=\dot{E}(r,\theta,\varphi,t)$.  The algorithm computes this quantity at  a set of user-defined
orbital phases $t_{i}$  within several orbital cycles that are also chosen by the user.  Because MESA performs
only a one-dimensional structure calculation, the spatial distribution of the tidal perturbations is not 
relevant in this paper.  Thus, we use the integral of $\dot{E}_\mathrm{V}$ over latitude and longitude for 
the energy dissipation rate profile in the radial direction: 

\begin{equation}
\dot{E}(r,t)=\int_0^\pi\int_0^{2\pi} \dot{E}(r,\theta,\varphi,t)~d\varphi~d\theta
\label{eq_dotErt}
\end{equation}

\noindent This quantity is the energy dissipation rate in the shell that lies at a distance $r$ from the
center el the star and has a total width $\Delta R$. Thus, it is the dissipated energy within a volume
$(4/3)\pi r^2 \Delta R$.  For the calculations in this paper $\Delta R=0.06\,R_1$.  Hence, the volume
of each shell is $\mbox{Vol}(r)=0.08\pi r^2 R_1$.

In general, we chose to output data at $N_\mathrm{tt}=40$ orbital phases within each of five orbital cycles.
An inspection of the five cycles allows assessment of the long-term variability patterns and
whether the initial transitory state has ended.  After the transitory state, the average angular velocities
evolve very slowly as the star tries to synchronize its rotation rate with the orbital angular velocity.
For the current MESA calculations we chose to use the last orbital cycle of the calculation and averaged
$\dot{E}(r,t)$ over the 40 orbital phases of the cycle to obtain the radial energy dissipation rate
profile, $\dot{E}(r)$, to be used as input for the MESA models: 

\begin{equation}
\dot{E}(r)=\frac{1}{N_\mathrm{tt}}\Sigma_{i=1}^{i=N_\mathrm{tt}}\dot{E}(r,t_i)
\label{eq_dotEr}
\end{equation}

\noindent Also of interest is the the total energy dissipation rate of the model:

\begin{equation}
\dot{E}_\mathrm{tot}=\int_{r_\mathrm{min}}^{R_1} \dot{E}(r)\,dr
\label{eq_dotEtot}
\end{equation}

\noindent where $r_\mathrm{min}$ is the radius of the shell that is contiguous to the rigidly rotating core.

\begin{deluxetable}{clcccc}
\tablecaption{Tidal perturbations models \label{tab:cases_run}}
\tablecolumns{6}
\tablewidth{0pt}
\tablehead{\colhead{Case} & \colhead{$m_2$}  & \colhead{$R_1$} &\colhead{$\beta_0$} & 
\colhead{$\dot{E}_\mathrm{tot}^{(\mathrm{a})}$} &\colhead{$\dot{E}_\mathrm{tot}$/$L_\odot$} }
\startdata
1b      & 0.8    &  1.20  &      0.7  &10.0 & 2.6 \\
1c      & 0.8    &  1.32  &      0.7  &24.1 & 6.3  \\
1d      & 0.8    &  1.50  &      0.7  &82.2 & 21.3 \\
2b      & 0.4    &  1.20  &      0.7  &2.9  & 0.75 \\
2c      & 0.4    &  1.32  &      0.7  &8.0  & 2.1 \\
2d      & 0.4    &  1.50  &      0.7  &26.8 & 7.0 \\
3b      & 0.4    &  1.20  &      1.1  &0.1  & 0.03 \\
3c      & 0.4    &  1.32  &      1.1  &0.3  & 0.08 \\
3d      & 0.4    &  1.50  &      1.1  &0.9  & 0.23 \\
\enddata
\tablenotetext{}{(a) Total energy dissipation rate, given in units of $10^{33}$~erg\,s$^{-1}$, is 
the average over time for the 50th orbital cycle considering both hemispheres of the star.
Column~6 gives this value in solar units.}
\end{deluxetable}

\begin{figure}
\includegraphics[width=0.95\columnwidth]{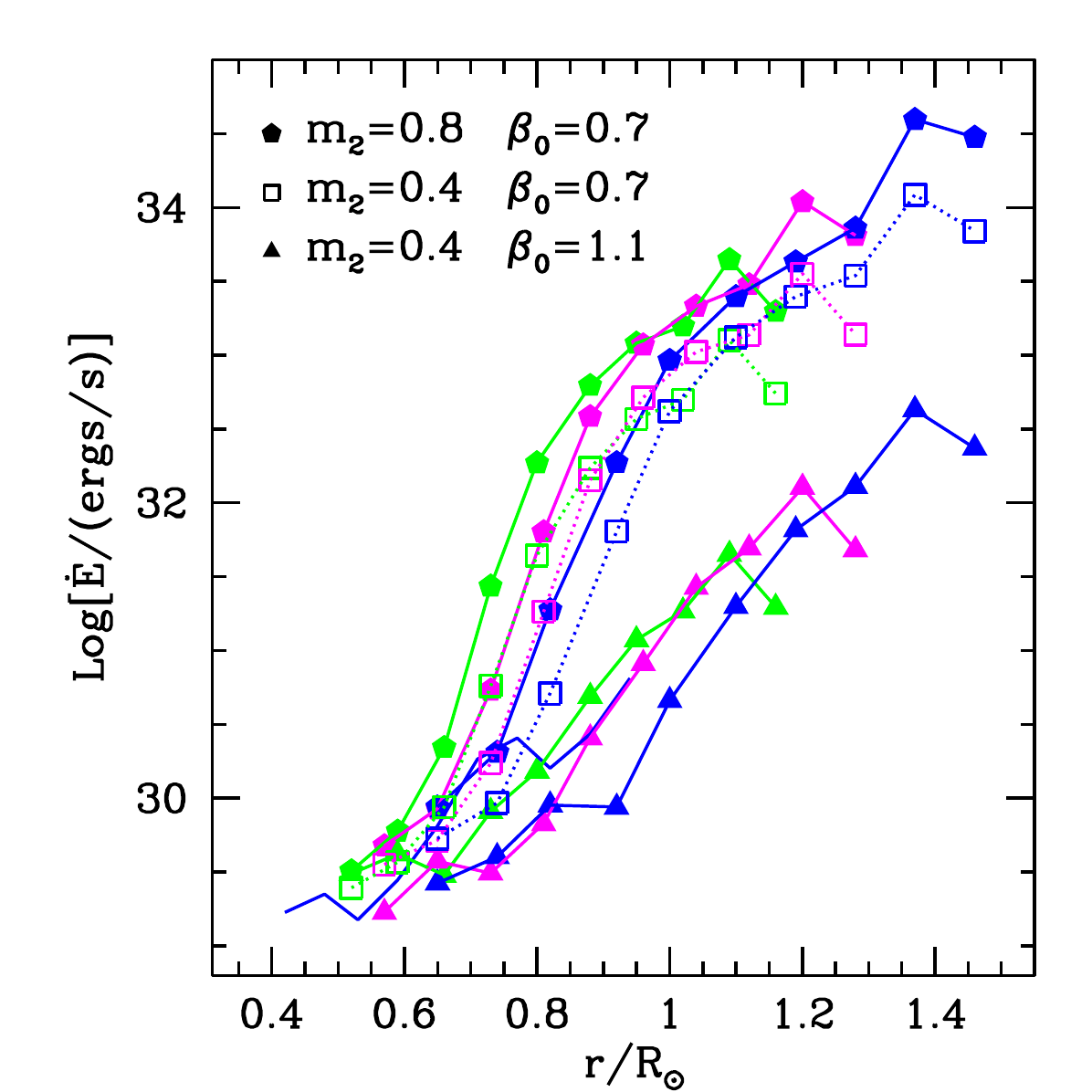}
\caption{Radial profile of the tidal energy dissipation rate as a function of distance from the
star in solar units. Hexagons show results for secondary mass $m_2=0.8\,M_\odot$
and $\beta_0=0.7$, squares for $m_2=0.4\,M_\odot$, $\beta_0=0.7$, and triangles for $m_2=0.4\,M_\odot$,
$\beta_0=1.1$. The colors indicate the radius: $R_1/R_\odot=1.20, 1.32, 1.50$ (green, magenta and blue 
respectively).  This figure shows that the $\beta_0$=0.7 model with the $m_2=0.4~M_\odot$ 
companion has larger dissipation rates than the $\beta_0$=1.1 model due to its larger departure 
from synchronicity, despite the slower rotation velocity.
}
\label{dotE_vs_radius}
\end{figure}

Figure~\ref{dotE_vs_radius} shows that $\dot{E}(r)$ displays a pronounced increase between the
lower boundary of the calculation (0.4\,$R_\odot$) and the surface. The rate of increase depends 
primarily on the degree of departure from synchronicity,  $|\beta_0-1|$, and less so on the mass 
of the companion.  This can be seen by comparing Cases~1 and 2 ($m_2=0.8$ and $m_2=0.4$, respectively), 
which differ in $\dot{E}(r)$ by a factor $\sim3$, while Cases~2 and 3 (both with $m_2=0.4$, 
but with $\beta_0=0.7$ and $\beta_0=1.1$, respectively) differ by $\sim$two orders of magnitude.  
Also noteworthy is the fact that $\dot{E}(r)$ is larger in the $\beta_0=0.7$ model than the corresponding 
$\beta_0=1.1$ model even though the former is rotating  more slowly than the latter;  for example, 
$\vv_\mathrm{rot}= 32$\,km\,s$^{-1}$ for $\beta_0$=0.7 while $\vv_\mathrm{rot}=51$\,km\,s$^{-1}$ 
for $\beta_0=1.1$.

There are several factors that can affect the energy dissipation rates obtained for any given model.
The model assumes that the central part of the star, below the smallest radius adopted for the
TIDES calculation, is in solid body rotation.  Even though the tidal perturbations in this region
induce negligible velocity perturbations,  the possibility exists that it has an intrinsic differential
rotation. This would add  shear energy dissipation to these inner zones and potentially affect the
stellar structure as described in the next section.  However, asteroseismological
studies of pulsating stars in the mass-range 1.4-5 $M_\odot$ have led to the conclusion that, for 
rotation velocities up to 50\% of breakup, the stars deviate from solid body rotation only mildly
\citep{2017ApJ...847L...7A}.  Another factor concerns the viscosity.  Our calculations were performed 
with $\lambda=1$, which most likely is an overestimate. However, the fact that $\dot{E}_\mathrm{V}$ 
scales approximately linearly with $\nu_\mathrm{turb}$ which, in turn, scales linearly
with $\lambda$ in the above prescription, allows scaling the values for smaller $\lambda$
values. 

\subsection{MESA models}\label{sec:MESA}

We use the open-source stellar structure and evolution code MESA,
version 15140\footnote{\url{https://docs.mesastar.org/en/r15140/}},
\citep{{2011ApJS..192....3P},
  {2013ApJS..208....4P},{2015ApJS..220...15P},{2018ApJS..234...34P},{2019ApJS..243...10P}}
to study the effect of tidal heating on the stellar envelope and resulting
observable properties. The basic model physics parameters (i.e.,
abundances, equation of state, opacities, reaction rates, diffusion,
convection, overshooting, semiconvection, thermohaline, boundary
conditions) are the same as those used by the MIST project \citep[see
  Table~1 of ][]{Choi2016} for low-mass stars, which were calibrated
using the old open cluster M67. We made the adjustments necessary to update the input file commands
from MESA v7503 used by MIST \citep{Dotter2016,Choi2016} to v15140
used in this paper. We do not include rotation in our models.

Stellar models were run for low-mass stars with masses close to the
turn-off mass (around 1.26\,$M_\odot$) for M67, which we use as a
reference and source of observational comparisons. The energy
dissipation rates as a function of radius, $\dot{E}(r)$, calculated by TIDES are 
used as inputs to the \verb|other_energy| hook in the MESA
code and allow us to calculate the effect of energy dissipation on the
stellar envelope of the model stars. In particular, as our fiducial
energy input, we chose the Case~1 energy dissipation profile 
(i.e., $m_2 = 0.8\,M_\odot$, $\beta_0 = 0.7$), which is shown in
Fig.~\ref{dotE_vs_radius} and listed in Table~\ref{table_ptk35_b} (see Appendix).

The TIDES tidal perturbation model is not an evolution code and the
outer radius of a given model is fixed. TIDES models with different
radii can therefore be taken to represent possible points along the
evolutionary track of a star. At any given time, the radius of a MESA
stellar evolution model can be compared with the list of TIDES model
radii. Interpolation between TIDES models of different radii enables
us to find the appropriate energy dissipation rates for each MESA
model in an evolutionary sequence. The TIDES shell radial positions
are at fixed fractions of the outer radius of the star, and so this
interpolation is performed using the normalized TIDES shell radii. The
mass of a given shell is the same between TIDES models and can be
easily found from the appropriate polytropic models. These masses are
used to calculate the final energy dissipation profile per unit mass,
which is then mapped back onto the MESA grid. MESA uses a much finer
grid of interior points than TIDES, and so linear interpolation is
used to provide values for the energy dissipation rate per unit mass
at intermediate positions. The heating profiles for MESA models that
have grown beyond the radius of the largest TIDES model are taken to be
that of $R_1 = 1.5\,R_\odot$.

As a reference, we compute the evolution of a $1.2\,M_\odot$ star with
no heating using the MIST set-up detailed in \citet{Choi2016}. This
model, which we refer to as the ``standard model'', was first evolved from
the birthline to the zero-age main sequence (ZAMS), which it reaches
after $2.6\times10^7$~yr. The ZAMS model was
then used as the starting point for all other models with the same
stellar mass. The standard model was evolved for 5\,Gyr, by which
time the star has left the main sequence and has entered the subgiant
stage. This is far longer than the age of M67, which is estimated to
be around 4\,Gyr, but allows us to study the effect of envelope heating
on more than one stage of evolution of the same star.

We examined two general scenarios:

\begin{enumerate}

\item{Scenario 1:}  Tidal heating is present throughout the main sequence.  This scenario
is possible only if the main sequence lifetime is on the order of the synchronization timescale.
Given the long lifetime of stars in the mass range we analyze here, this will not occur unless
the binary is forced to remain in an eccentric orbit by external gravitational forces~--~for example
a third component in the system.

\item{Scenario 2:} Tidal heating occurs in a short burst ($10^7$~yr)
  at a time when the internal structure undergoes a relatively
fast transition, such as when the primary star reaches the end of its core
hydrogen-burning phase, and the binary may become asynchronous.
 
\end{enumerate}

For Scenario 1, several tests were performed and the resulting models are labeled as follows: 
\begin{itemize}
\item{} h00: this is the standard model. The star evolves from the ZAMS to the stopping point with no 
added heat;
\item{} h10: after age $10^8$~yr heat is injected steadily using the radial profile of the energy dissipation rate for
the 1.2$+$0.8\,$M_\odot$ system shown in Fig.~\ref{dotE_vs_radius}.

\item{} h05 and h01: same as h10 but only 50\%  and 10\%, respectively,
  of the radial energy dissipation rates are injected into the
MESA model.  These tests provide insight into the range of heating rates that have a
significant effect on the stellar structure.


\item{} hk10: same as h10 except that the heat is not injected into
  the outermost 10\% of the star.  The reason
for this test is that very near the surface, where the optical depth
$\tau\leq 2/3$, \citet{1975A&A....41..329Z}
suggested that a significant amount of the excess energy could be rapidly radiated away.  Also, this
test serves to examine the impact of the large heating rates in the  surface layers on the internal
structure.

\end{itemize}

Scenario 2 is investigated for stars with 1.16, 1.2, 1.26, 1.29 and
1.32\,$M_\odot$ with the aim of making direct comparisons with observed
stars in a range of evolutionary states in the M67 open cluster.  In this scenario,
MESA stellar evolution models for each mass were
run without any heating until an age of 3.97~Gyr at which time the h10 heating rate was injected for
a timespan of $10^7$~yr. This timescale is so short compared to the stellar lifetime that it can 
be viewed as a heating pulse, as we refer to it in what follows.

For the heating pulse,  we set up the calculation  to have a maximum
timestep of $10^5$~yr, which is much finer than our standard MESA
model runs, in order to capture the details of this transient episode
in the life of the star.  The computations were halted at an age of 4.25~Gyr but
the main focus is around 3.98~Gyr, which the MIST isochrone shows to
be a good fit to the observed stars in M67. 

~ ~  ~  ~  
\begin{deluxetable}{llllllll}
\tablecaption{MESA models: Ages and luminosities of the models with continuous heating at radii  
$R_1/R_\odot=1.20$, 1.32, 1.50 (see Fig.~\ref{structure_R1.5}).   
\label{ages_radii}}
\tablecolumns{8}
\tablewidth{0pt}
\tablehead{\colhead{Heating} &\colhead{Name} &\multicolumn{3}{c}{Age/Gyr}& \multicolumn{3}{c}{Luminosity/$L_\odot$}\\
\hline
                &    &    \colhead{1.20}&\colhead{1.32}&\colhead{1.50} & \colhead{1.20}&\colhead{1.32}&\colhead{1.50}
}
\startdata
None  & h00   &1.35   &2.75     &  4.10  & 2.22    &  2.66       &  3.01   \\
0.1   & h01   &1.25   &2.60     &  3.80  & 2.24    &  2.70       &  3.29   \\
0.5   & h05   &0.90   &2.25     &  3.35  & 2.29    &  2.85       &  3.65   \\
1.0   & h10   &0.50   &1.90     &  3.10  & 2.33    &  2.92       &  3.83   \\
1.0   & hk10   &0.70   &2.25     & 3.45  & 2.31    &  2.86       &  3.54    \\
\enddata
\end{deluxetable}

\section{Results} \label{sec:results}

\subsection{Scenario 1: Continuous heating} \label{sec:stellar_structure}

\begin{figure}
\includegraphics[width=\columnwidth]{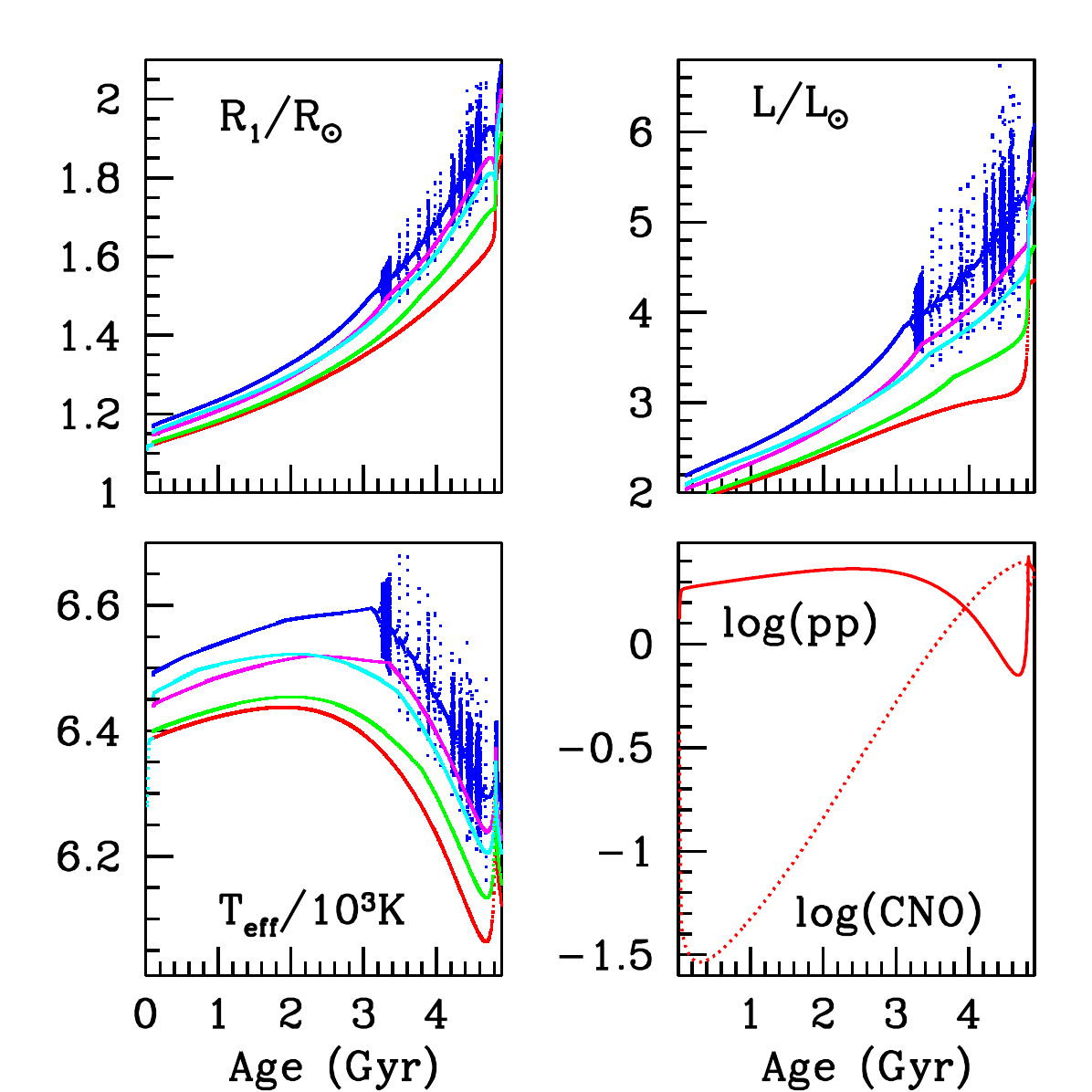}
\caption{Properties of the MESA models with continuous heating.  Stellar radius (top left), effective
temperature  (bottom left) and surface luminosity (top right). Core hydrogen fusion energy generation   
rates are shown in the bottom right panel for the proton-proton reaction chain (solid)  and the CNO 
reaction chain (dots) only for the standard model.  The colors correspond to the standard 
model (red) and the heated models h10 (blue), h05 (magenta), h01 (green), and hk10 (cyan).
Model~h10 develops instabilities starting at an age $\sim$3\,Gyr, but similar instabilities
are not present in model hk10 in which the heating was injected only
in layers  having $r \leq 0.85 R_1$,
where $R_1$ is the surface radius.
}
\label{parameters_age}
\end{figure}

Figure~\ref{parameters_age} shows the manner in which the tidal heating affects the basic parameters
of the perturbed star compared to the standard model.  
The heated star grows to a larger radius and this bloating scales with the amount of injected energy.  
Similarly, the effective temperature and surface luminosity are also larger than in the standard model.  
This description applies to all ages $\leq$5~Gyr, which is when the
MESA model was terminated. The model with the highest heating rate
(model h10) develops oscillations in radius, effective temperature and
luminosity. The onset occurs around 3~Gyr, which is when the stellar radius
has exceeded 1.5\,$R_\odot$ and the heating rate becomes saturated at
the limit of the largest TIDES model.
Such instabilities are not present in model hk10, in which tidal heating is
injected only in layers below the surface, $r\leq 0.85 R_1$, where $R_1$ is the
surface radius. This case could apply to a star that has already
synchronized the outer layers with the binary orbit but not those below.

\begin{figure}
\includegraphics[width=\columnwidth]{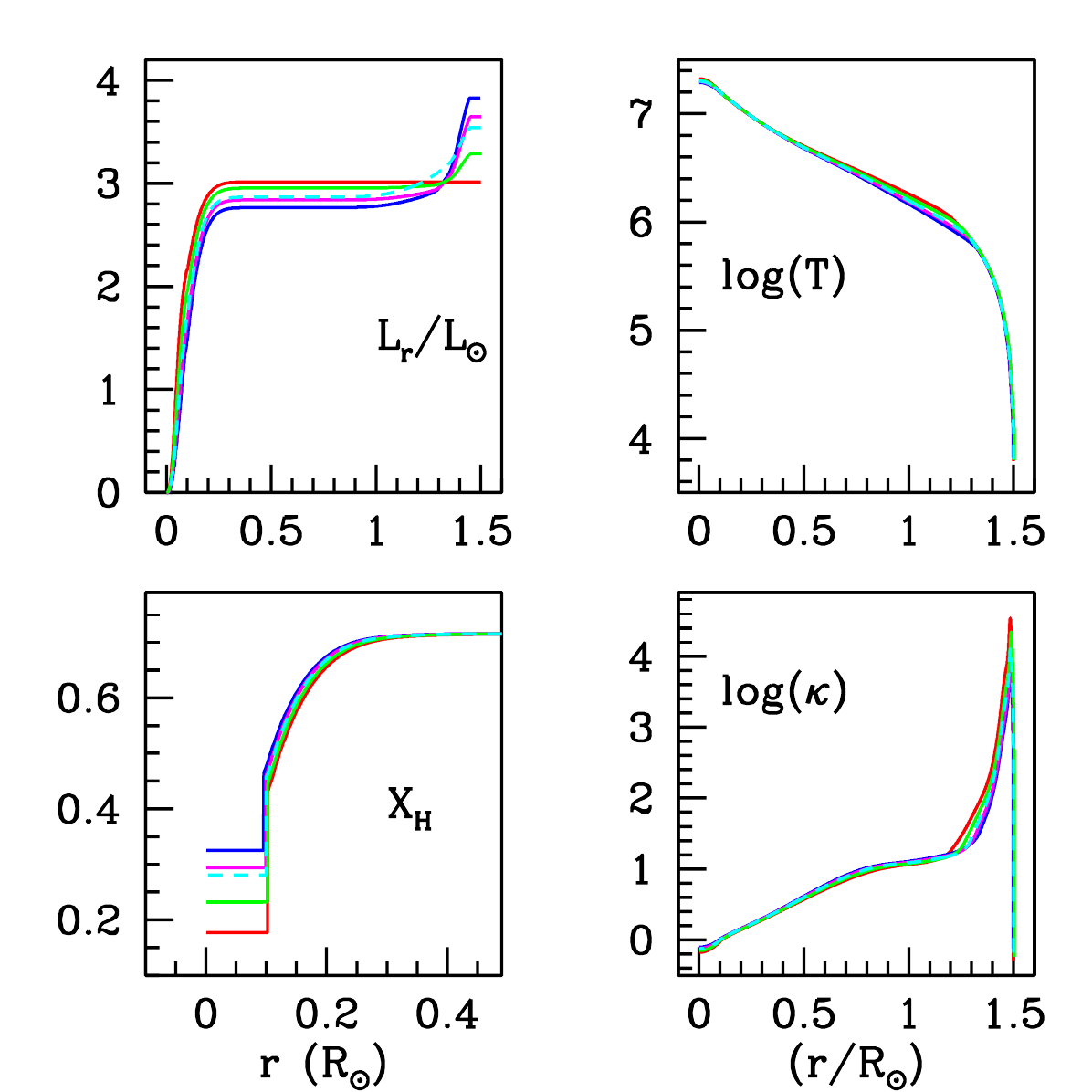}
\caption{Internal structure of the MESA models that were continuously heated, shown at the 
age at which the surface radius of each model has reached is $R_1=1.5\,R_\odot$
(see Table~\ref{ages_radii}). Red is the standard model (h00) at an age 4.1~Gyr, blue is 
the fully heated model (h10) at an age 3.1~Gyr, green is the model with 10\% the heating rate 
of h10 at an age 3.8~Gyr, and magenta is the model with 50\% the heating rate of h10 at an
age 3.35~Gyr.  Luminosity, $\log\left(L_r/L_\odot\right)$ (top left), hydrogen
mass fraction $X_\mathrm{H}$ (bottom left), 
temperature, $\log(T)$ (top right), and opacity, $\log\left(\kappa/\mbox{cm$^2$g$^{-1}$}\right)$ (bottom right).
}
\label{structure_R1.5}
\end{figure}

From an observational standpoint, eclipsing binaries are routinely used to determine the mass and
radius of the two stars which, in turn, are used to constrain the age of the system.  However,
if tidal heating is active, stars of a given radius can have different ages.  For example,
a star evolving according to the standard model attains a radius $R_1=1.5\,R_\odot$ at an age of
4.1\,Gyr while our h10 heated model attains this radius at an age of 3.1\,Gyr (see Table~\ref{ages_radii}).  
The internal structure of stars with the same radius but different heating rates will also be different.  
For example, Figure~\ref{structure_R1.5} presents a snapshot of the internal structure of the models 
at the point where the radius has expanded to 1.5~$R_\odot$.  As was seen in \citet{2023A&A...670A..44E}, 
the internal luminosity of heated models is lower than that of the standard model for $r\leq1.3\,R_\odot$ 
because the heated stars are younger. This is because the luminosity leaving the core increases with 
decreasing central hydrogen mass fraction, i.e.\ due to chemical abundance changes with age as a
result of fusion reactions. On the other hand, the luminosity of the heated models rises rapidly in 
the outer layers because of the injected heat, while that of the standard model remains constant between 
the edge of the core and the stellar surface.  The internal temperature and opacity do not change significantly.

\begin{figure}
\includegraphics[width=0.98\columnwidth]{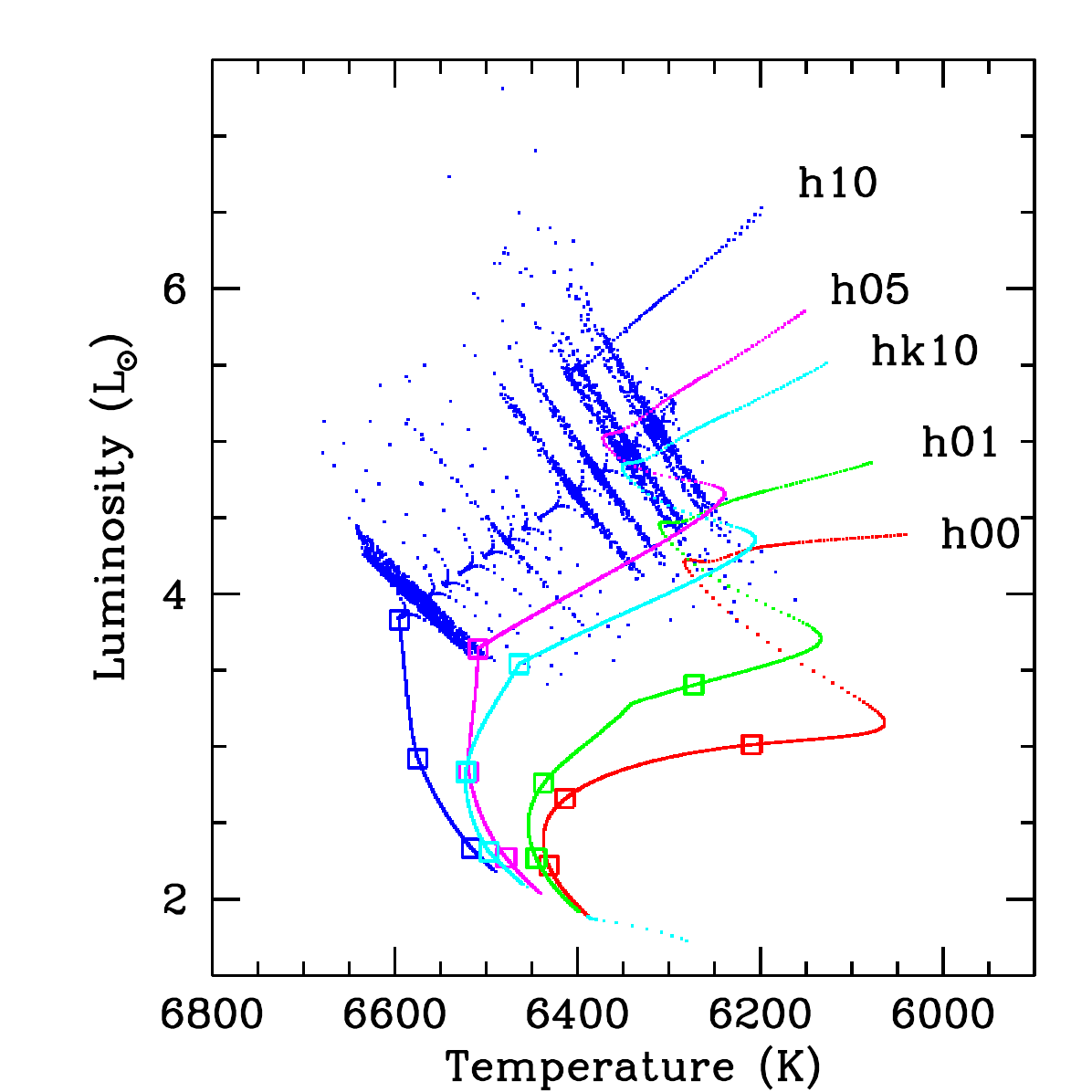}
\caption{Evolutionary tracks of the four MESA models computed with continuous heating: the 
standard model with no heating (h00, red), and heated models with, respectively, 0.1 (h01, green), 
0.5 (h05, magenta), and 1.0 (h10, blue) the heating rates predicted by the TIDES models.   Each 
MESA timestep is indicated by a dot.  The h10 track undergoes instabilities starting at an 
age $\sim$3.1~Gyr. Squares indicate the position in each track at 
which the stellar radius is 1.2, 1.32 and 1.50\,$R_\odot$ (see Table~\protect\ref{ages_radii}). 
}
\label{theoretical_HRD}
\end{figure}

The evolutionary tracks of the tidally heated models are compared to
that of the standard model in Fig.~\ref{theoretical_HRD}. The heated
models are characterized by higher temperatures and luminosities than
the reference case.  The difference between heated and standard models
increases with increasing heating rates.  The largest heating rate
(model h10) presents large oscillations shortly after the star reaches
a radius 1.5\,$R_\odot$.  Since MESA dynamically adjusts the
integration timestep, these oscillations do not appear to be due to
numerical instabilities. Instead, we tentatively interpret the oscillations as the
outer layers of the star adjusting to the heating rate becoming
saturated in the limit of the largest TIDES model ($R_1 = 1.5\,R_\odot$). 
A dynamical treatment of this model, beyond the scope of
the present paper, would be able to ascertain whether such
oscillations could lead to mass loss.


We note that the excess luminosity of the MESA models is significantly smaller than the total 
energy dissipation rate in the shearing layers;  that is, $L_\mathrm{exc}=\alpha \dot{E}$,  
where $L_\mathrm{exc}=L_\mathrm{heated} - L_\mathrm{standard}$ is the difference between the 
MESA heated and standard models for the same radius and  $\alpha<1$ is the fraction of dissipated energy 
that escapes as radiation and contributes to the luminosity. Comparing the results of model h00 with 
the heated models we find that $\alpha \leq 10\%$.\footnote{For example, at a time when the stellar radius 
reaches  1.5~$R_\odot$, the luminosity of the MESA models are, respectively, 3.01~$L_\odot$ (standard model), 
3.29~$L_\odot$ (h01), 3.65~$L_\odot$ (h05), and 3.83~$L_\odot$ (h10) (see Table~\ref{ages_radii}). Thus, 
$L_\mathrm{exc}=0.28$, 0.64 and 0.82~$L_\odot$, respectively.  The tidal shear energy dissipation rates 
that were injected into the MESA model are  2.6, 6, and 21~$L_\odot$, respectively.}
The remaining $\sim$90\% of the shear energy goes into sustaining the star against self-gravity and
inflating it.  

\subsection{Scenario 2: Pulsed heat injection}

The evolutionary paths shown in Fig.~\ref{theoretical_HRD} correspond to continuous heating,
a rather unrealistic state because the  main sequence lifetime of solar-type stars is $>10^9$~yr, 
while synchronization and orbital circularization processes are believed to last $\sim10^7$--$10^8$~yr.  
Hence,  even if a star reaches the main sequence with non-synchronous rotation or an eccentric orbit, 
it is expected to attain an equilibrium state early in its main sequence lifetime.  A more
realistic situation is one in which the system departs from the equilibrium condition at some point
during its evolutionary trajectory.  A clear example is when a star nears the end of core hydrogen 
burning and the core contracts and spins up while the envelope expands and spins down.  At this stage, 
the rotation becomes asynchronous and a differential rotation structure is established, thus 
triggering a tidal heating event. This is described by our {\it Scenario~2}, the pulsed-heating scenario. 

\begin{figure}
\centering
\includegraphics[width=0.29\columnwidth]{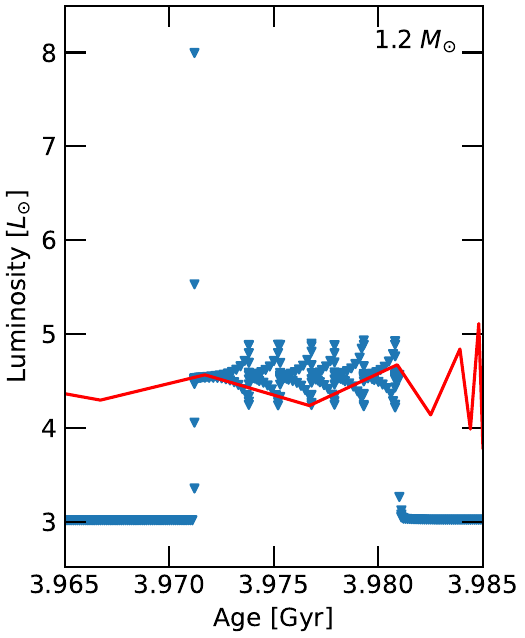}
\includegraphics[width=0.29\columnwidth]{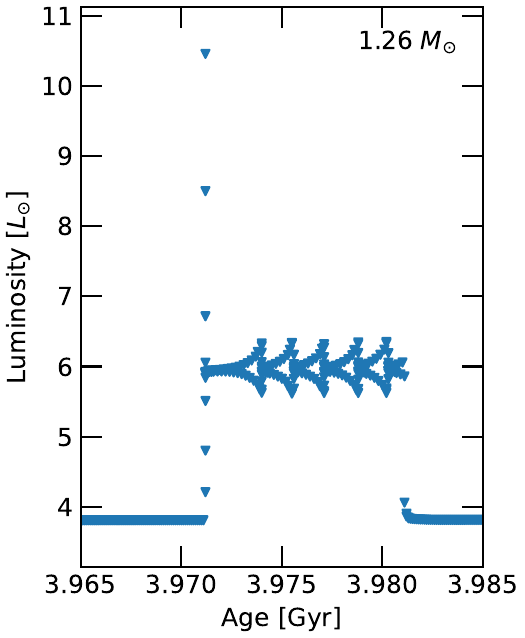}
\includegraphics[width=0.29\columnwidth]{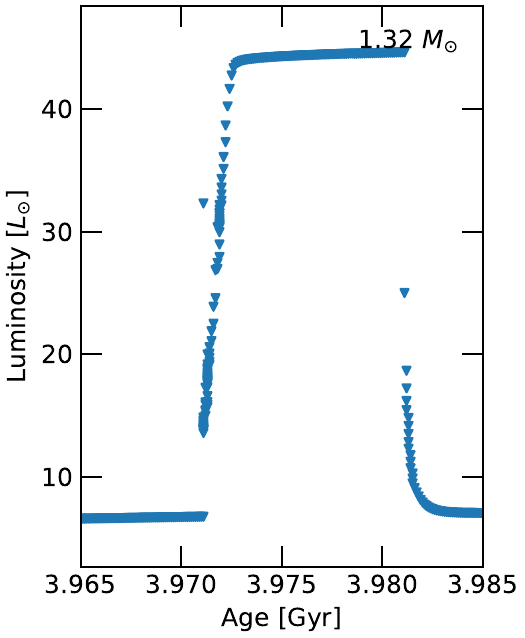}
\includegraphics[width=0.29\columnwidth]{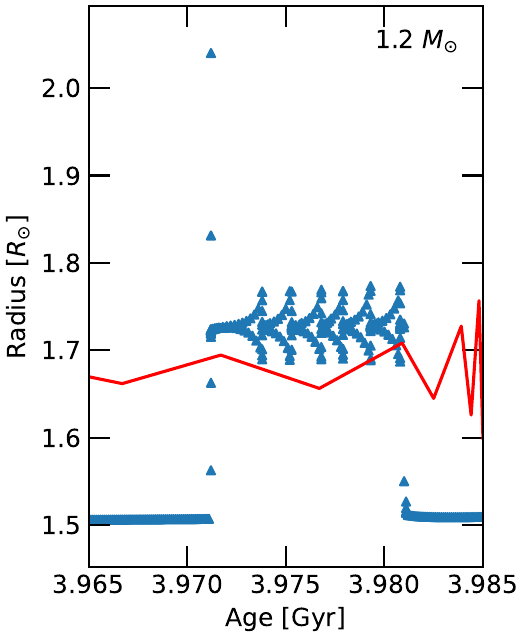}
\includegraphics[width=0.29\columnwidth]{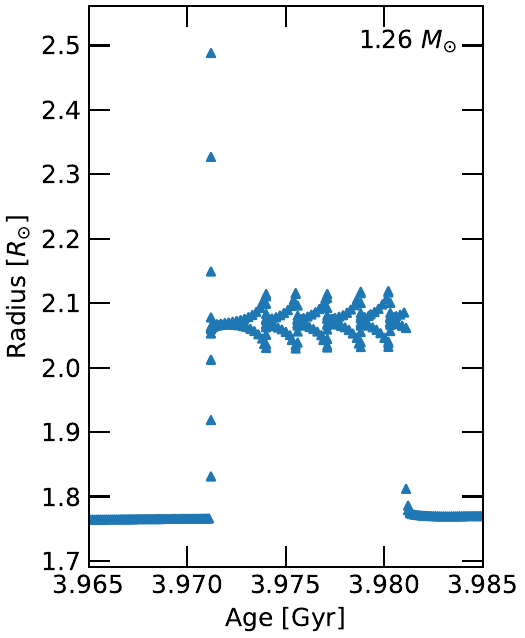}
\includegraphics[width=0.29\columnwidth]{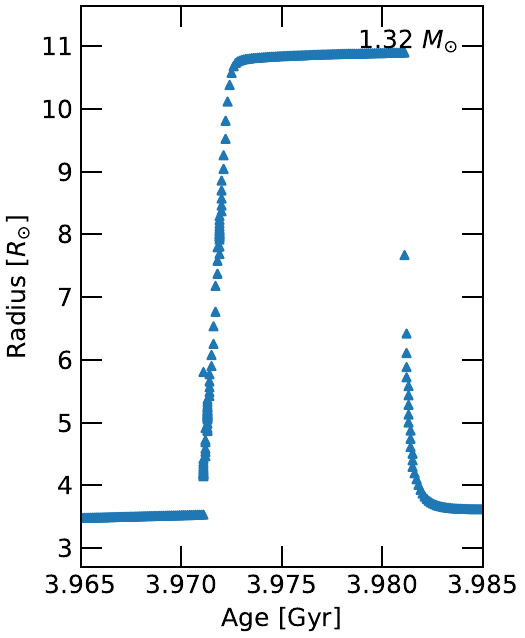}
\caption{Luminosity (top) and radius (bottom) against time of three MESA models with an energy injection 
pulse lasting $10^7$~yr. Examples for stars with masses 1.2, 1.26 and 1.32~$M_\odot$ are illustrated.  
The pulse starts at an age $\sim$3.97~Gyr, which is close to the main-sequence turnoff age of the 
M67 cluster. The red lines in the first column correspond to the continuous heating model h10 for 
the 1.2~$M_\odot$ star during this time interval.
}
\label{age_lum_radius_pulse}
\end{figure}

The evolution over time of luminosity and radius of the pulse-heated models are summarized in 
Fig.~\ref{age_lum_radius_pulse}. As soon as the heat is injected, the stellar response is a radius
and luminosity increase.  The response time of the 1.20 and 1.26~$M_\odot$ models is $\sim10^{5}$~yr, with
the luminosity and radius first increasing and then decreasing to an intermediate level where they
settle at a plateau.  At this time  they undergo a series of oscillations about the mean level. As soon 
as the heating is turned off, the luminosity and radius decrease to that of the standard model on a 
timescale that is as short as when the heat was first injected.   For the 1.20 and $1.26 M_\odot$ models, 
the mean luminosity and radius increases are modest, reaching a factor 1.5 increase in luminosity and 
only a factor 1.15 in radius. Also shown in Fig.~\ref{age_lum_radius_pulse} is the response of the 
$1.32 M_\odot$ model, which shares similar features with the lower-mass models but without the
oscillations. In addition, in this higher mass model, the luminosity increases by a factor $\sim 7.7$, 
while the radius grows by a factor $\sim 3.8$ over the $10^7$ yr duration of the heat injection.

Also shown in Fig.~\ref{age_lum_radius_pulse} is the response of the 1.32~$M_\odot$ model, which shares
similar features with the lower-mass models but, instead of remaining at a plateau with oscillations, it
continues brightening slowly until the heating is turned off.  The overall
luminosity increase is nearly a factor of three,  and the radius increase mimics that of the luminosity, 
nearly doubling in size over the $10^7$~yr duration of the heat injection.

\begin{figure}
\includegraphics[width=0.95\columnwidth]{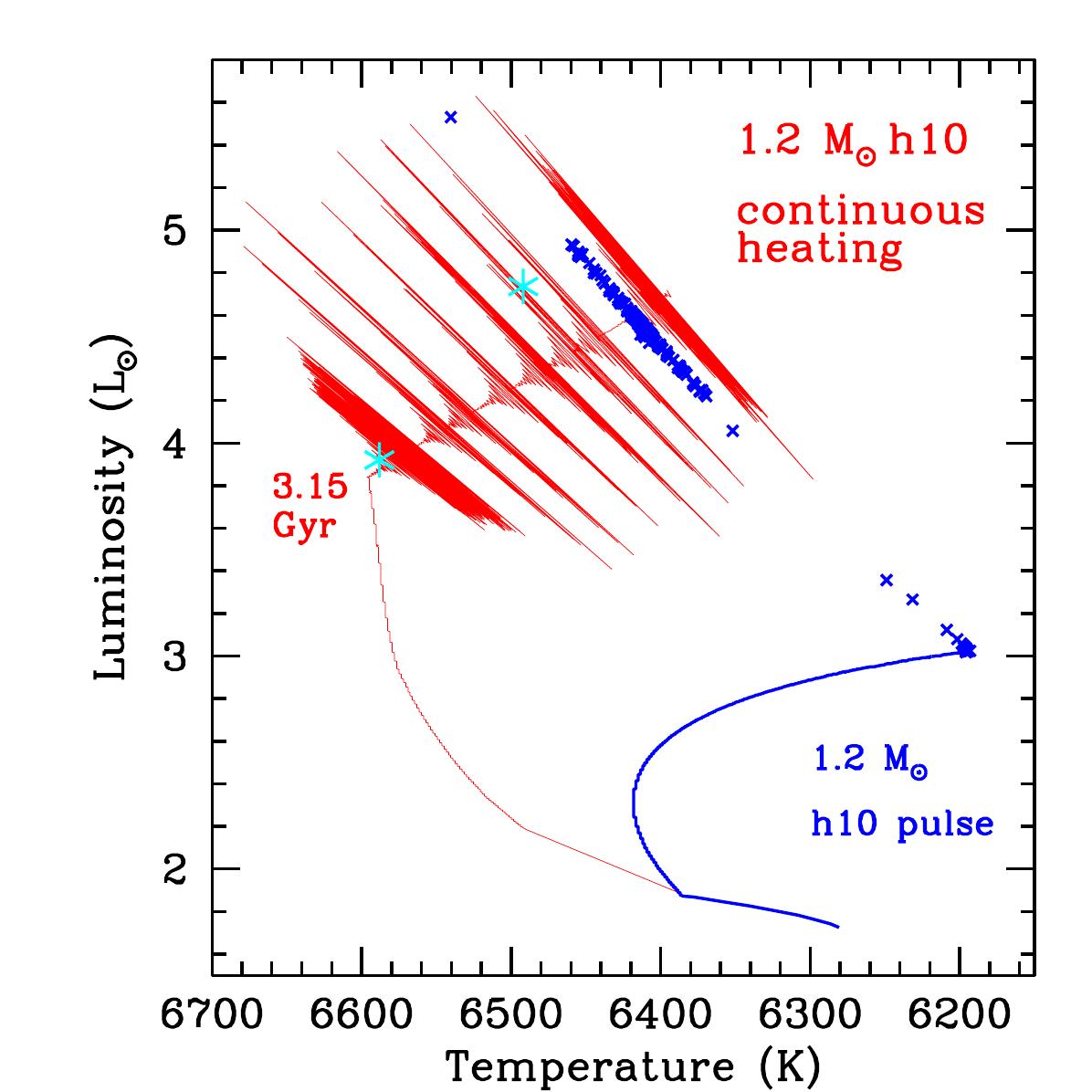}  
\caption{Evolutionary track for the 1.2~M$_\odot$ model with a pulse of heat
injection (blue) at 3.98 Gyr compared to the model with continuous heating (red).  The
pulse-heated star reaches the same location on the HRD  as the continuously heated models. 
The crosses mark the time during which the heat was injected. The asterisks mark the locations
on the continuously heated trajectory corresponding to 3.15 Gyr and 3.98 Gyr.
}
\label{compare_pulse_m2}
\end{figure}

\citet{2023A&A...670A..44E} found that the evolutionary track of a star in which the
tidal heating is introduced late during its main sequence lifetime rapidly reaches
the evolutionary track of a star with continuous heating throughout the main sequence.  
We confirm this result in our current calculations.  Figure~\ref{compare_pulse_m2}  shows that
the evolutionary track of the pulse-heated model intersects that of of the h10 continuously 
heated model during the duration of the pulse. Thus, the continuously heated track serves to 
represent the locus of locations on the HRD that will be occupied during a tidal heating event.  
The precise location will depend on the age at which the heating event is triggered.

\begin{figure}
\includegraphics[width=0.95\columnwidth]{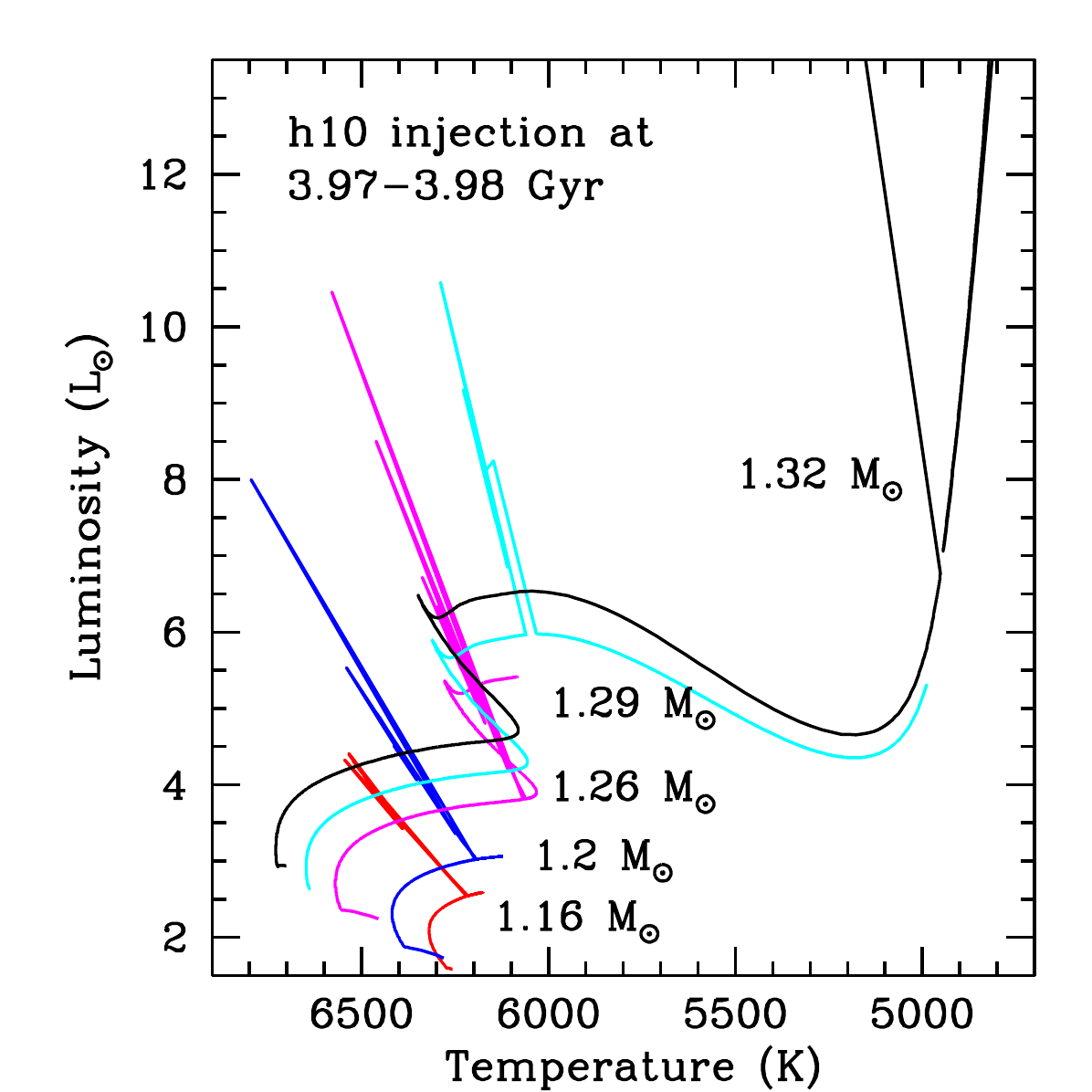}
\caption{Evolutionary tracks with a pulsed heating episode at an age of 3.98~Gyr
for $M/M_\odot= 1.16$ (blue), 1.2 (red), 1.26 (magenta), 1.29 (cyan) and 1.32 (black).
Each track ends at an age of 4.25~Gyr.  During the heating pulse phase,
the tracks of the lower-mass stars intersect tracks of more massive stars. 
The track for $1.32 M_\odot$ is truncated and is shown separately in Fig.~\ref{theoretical_pulsed_1m32}.
}
\label{theoretical_pulsed}
\end{figure}

The theoretical HRD for masses in the range 1.16\,$M_\odot$ and 1.29\,$M_\odot$ with heat 
injection starting at an age of 3.97\,Gyr and ending in 3.98\,Gyr is displayed in 
Fig.~\ref{theoretical_pulsed}.  In all cases, the star crosses through the standard tracks of 
more massive stars during the heat pulse.

Another important feature of the pulses is that as soon as the heating is turned off, the star rapidly
returns to the standard evolutionary track.  Thus, the permanence of a tidally heated star on 
the blue side of the standard evolutionary track is as brief as the the time during which
the heating is injected.  Considering the long lives of these types of stars, the heating
event can be considered to be extremely brief, thus explaining the relatively small number of stars that
would be in this state at any given time.

\begin{figure}
\includegraphics[width=0.95\columnwidth]{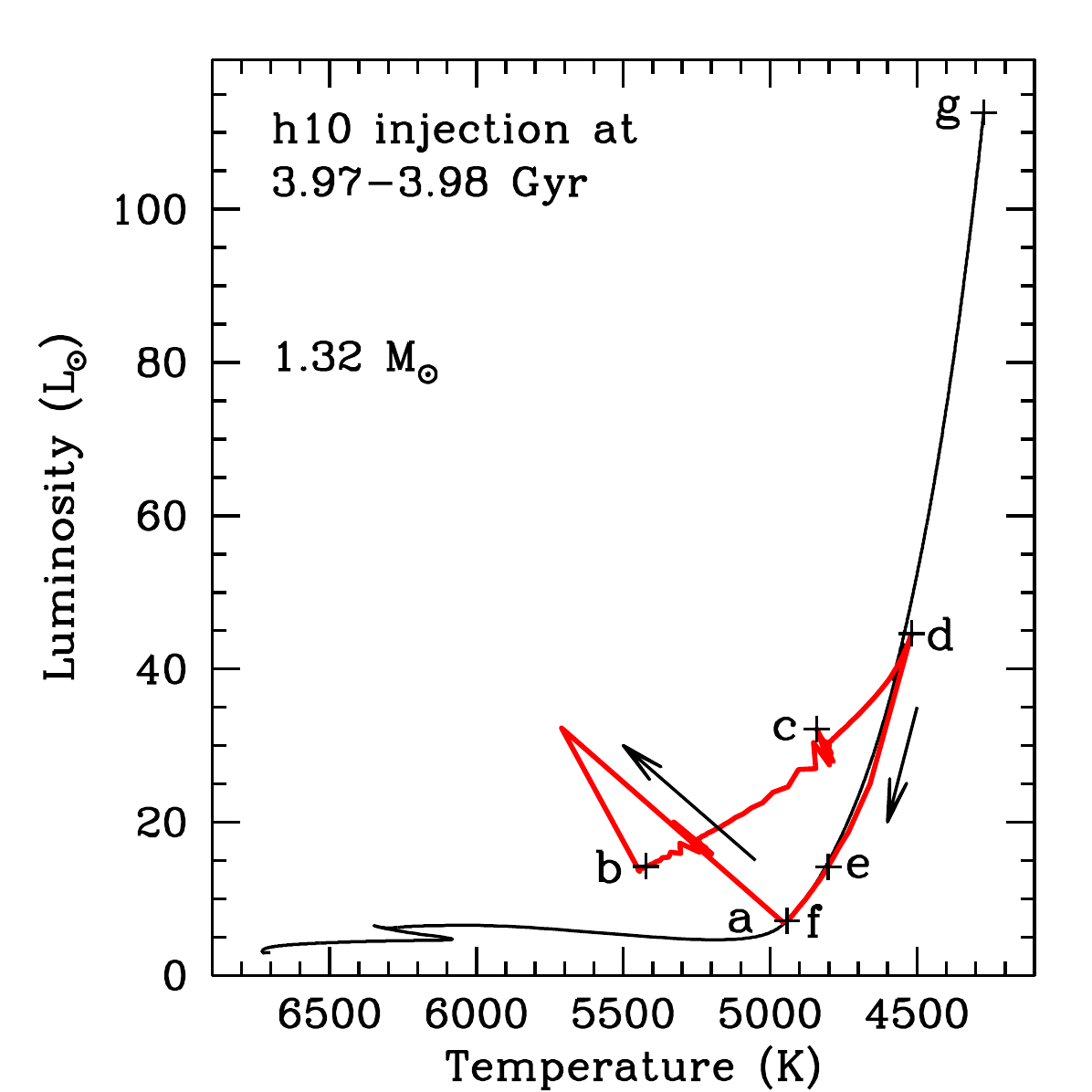}
\caption{Evolutionary track of a $1.32 M_\odot$ star with a pulsed heating event. 
Red color indicates the part of the HRD track that is affected by the heating event. 
The crosses mark a selection of points during and immediately after the heated trajectory. These
points are labeled with a letter indicating the corresponding age in Gyr as follows: (a) 3.970 
(heating is turned on); (b) 3.971; (c) 3.972; (d) 3.980 (heating is turned off); (e) 3.981; 
(f) 3.989 (coincides with a); (g) 4.25 (end of the calculation).  The arrows mark the direction 
of the trajectory  after heating is initiated and immediately after it is turned off. 
}
\label{theoretical_pulsed_1m32}
\end{figure}

Further insight into the timescales involved can be gained by a
detailed look at the 1.32~$M_\odot$ model undergoing a tidal heating
episode that lasts 10~Myr, as illustrated in
Fig.~\ref{theoretical_pulsed_1m32}.  The heat is first injected at an
age of 3.97~Gyr and causes an immediate, transitory spike in
luminosity and surface temperature, but the star then settles onto a
new trajectory higher up the HRD with increasing luminosity and
decreasing temperature. The first $\sim$1~Myr of this trajectory
exhibits oscillatory behaviour superimposed on the general trend and
by the end of this stage (point 3.9719 in the figure) the surface
temperature has returned to its original value but the luminosity is
four times higher. For the remainder of the heat-pulse duration,
$\sim8.9$~Myr, the star's luminosity continues to rise and the
temperature continues to decrease, until the trajectory intercepts the
red-giant branch shortly before the heating switches off (point
3.980). Once the heating is turned off, the star quickly descends the
red-giant branch, decreasing in luminosity while its surface
temperature remains roughly constant. It takes $\sim$8~Myr for the
star to return to its original location in the HRD, with the
luminosity decreasing from $45 L_\odot$ to $8 L_\odot$ in the first
$\sim$1~Myr but taking a further 7~Myr to fall to $7 L_\odot$. At this
point the star retakes the original trajectory and the luminosity
begins to increase again as the star climbs back up the red-giant
branch. The increases (decreases) in the luminosity due to the heat
pulse are accompanied by increases (decreases) in the stellar radius.

The appearance of oscillations in the heated 1.2 $M_\odot$ and 1.26 $M_\odot$ models merits
further comment. The layers into which the heat is injected lie very far from the nuclear processing
zone which, for the ages considered, is limited to the central region of the star. They do, however,
include the surface convection zone.   Kippenhahn diagrams of the models show that the oscillations
are present at the base of the convection zone, within the radiative zone (see Fig.~\ref{kipp_pulsed}).  
These oscillations are forced and for this reason only last the duration of the heat pulse. When 
heat is added to the outer layers of the model stars, these layers expand and their densities fall. 
This reduces the opacity in these layers (which correspond to temperatures $T<10^6$ K, hence the 
opacities are Kramers-type) and so the radiative region of the star now extends to larger radii. 
The partially ionized zones of H and He are also lifted to larger radii. Although physical 
oscillations may occur under such circumstances, we do not model them since our minimum time step 
is of order 1000 years, far longer than any physical oscillation. The oscillatory behaviour 
exhibited in Figure 5 is a result of fluctuations in the forcing term due to the mapping between 
the stellar model and the TIDES model. This affects the base of the heating region, since its 
position depends on the total radius of the star, and as a consequence all the layers above 
receive different heat forcing at each timestep, and in particular the base of the convection zone. 
The behaviour appears oscillatory because the mapped heating at each position diminishes as the 
stellar radius grows, i.e. the forcing becomes less and this leads to damping. Determining 
whether  true oscillations during a heating pulse can be excited requires a separate investigation
which is beyond the scope of this paper.

\section{Tidal heating and the isochrone of M67} \label{sec:M67}

\begin{figure}
\centering
\includegraphics[width=0.49\columnwidth]{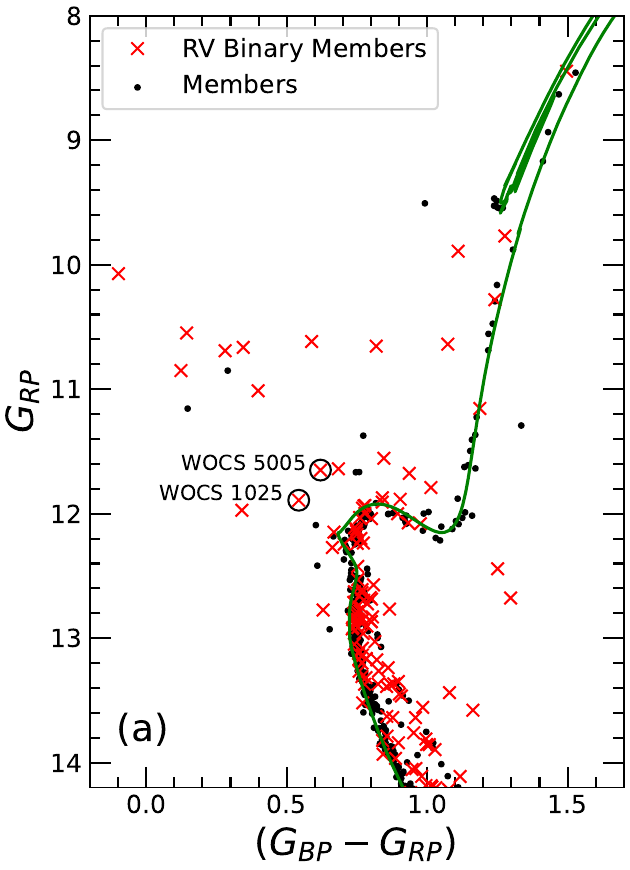}
\includegraphics[width=0.49\columnwidth]{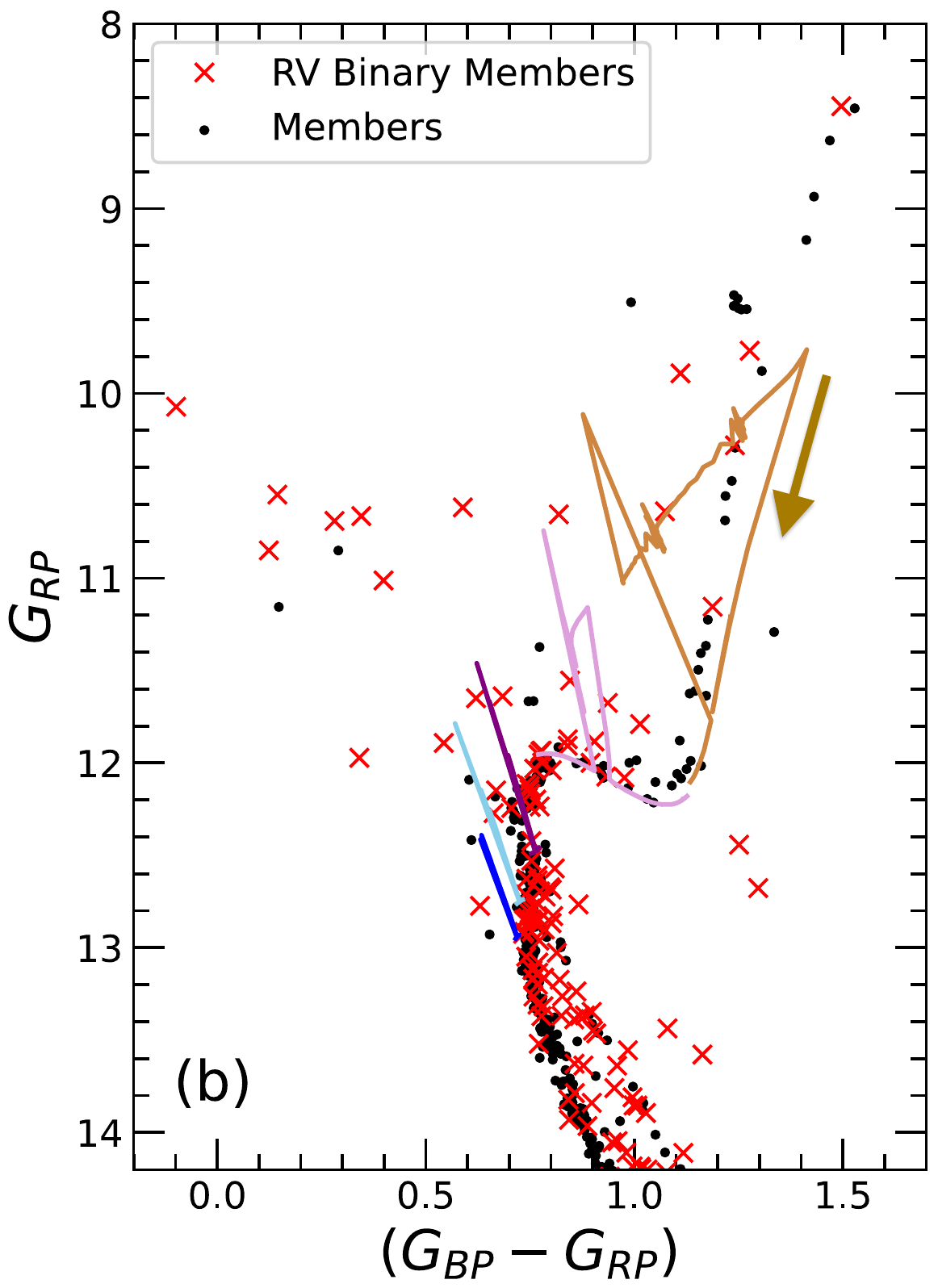}
\caption{Color-magnitude diagrams with \textit{Gaia} DR2 photometry showing M67 designated member stars from \citet{Geller2015} together with the MIST 3.98~Gyr isochrone and MESA model partial evolutionary tracks centred around the heat pulse time. (a) Isochrone (green solid line) and M67 designated binary (red crosses) and member stars (black dots). (b) M67 stars and partial evolutionary tracks: $1.16 M_\odot$ (blue line), $1.2 M_\odot$ (light blue line), $1.26 M_\odot$ (purple line), $1.3 M_\odot$ (pink line), $1.32 M_\odot$ (brown line). The brown arrow indicates the direction of the evolutionary track of the $1.32 M_\odot$ star after the heat pulse. 
}
\label{cmd-iso-pulse}
\end{figure}
\begin{figure}
\centering
\includegraphics[width=0.49\columnwidth]{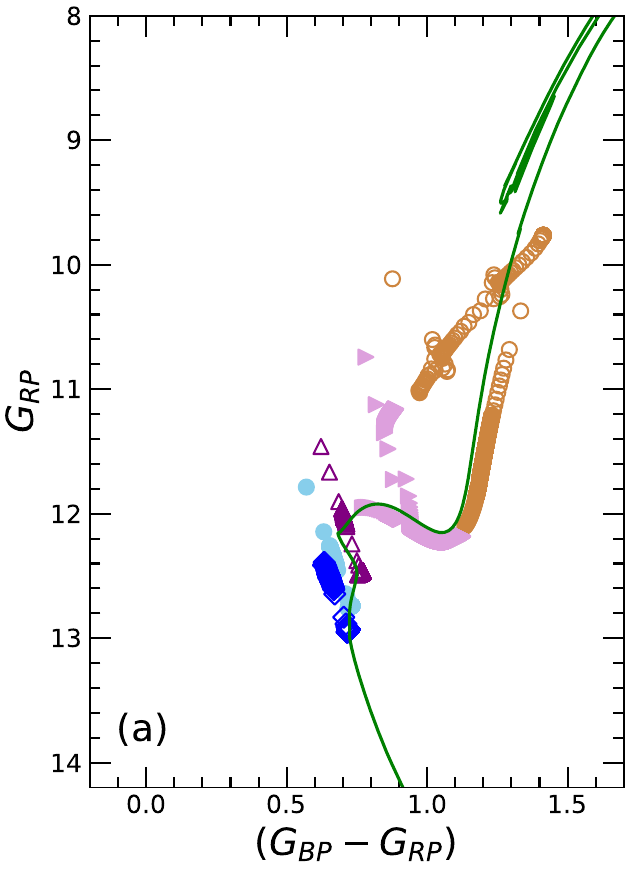}
\includegraphics[width=0.49\columnwidth]{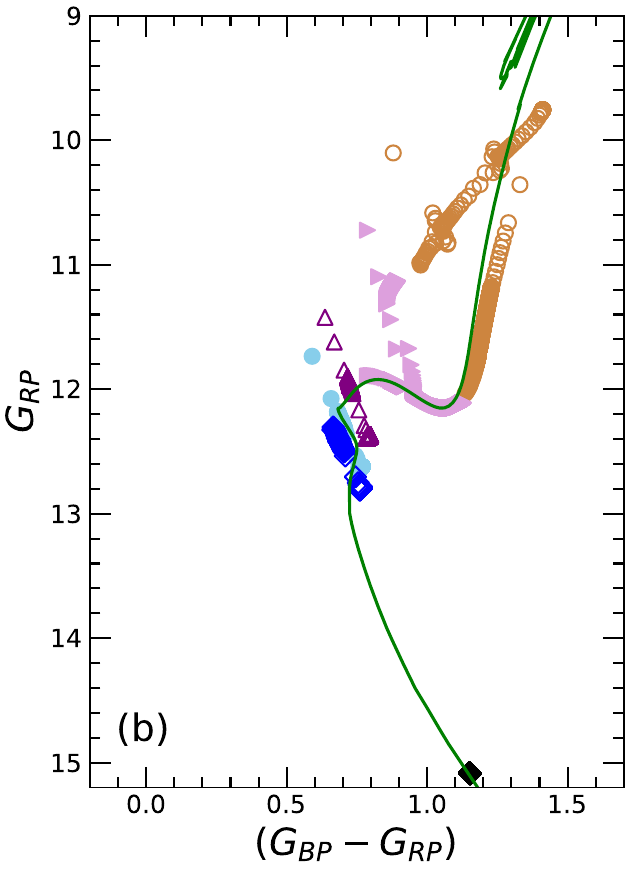}
\caption{Isochrone and partial evolutionary tracks: $1.16 M_\odot$ (blue open diamonds), $1.2 M_\odot$ (light blue filled circles), $1.26 M_\odot$ (purple open triangles), $1.3 M_\odot$ (pink filled triangles), $1.32 M_\odot$ (brown open circles). Left panel: single star case. Right panel: includes the contribution of the $0.8 M_\odot$ binary companion (black filled diamond) to each model star color and magnitude. Note that the magnitude axis has been shifted by 1~mag to show the position of the $0.8 M_\odot$ companion if it were an isolated star.
}
\label{cmd-bin-pulse}
\end{figure}

The old open cluster M67 has an age of around 4~Gyr, which has been determined by a variety of methods including white dwarf cooling \citep{Bellini2010}, isochrone fitting \citep[e.g.,][]{Sandquist2021,Nguyen2022} and gyrochronology \citep{Barnes2016}.
It is used as a case study for stellar evolution codes for testing internal stellar physics such as convective overshooting and diffusion \citep{Choi2016,Nguyen2022}. In particular, we obtained MIST isochrones for non-rotating models with metallicity $\mbox{[Z/H]}=0.0$ from the MIST database\footnote{\url{https://waps.cfa.harvard.edu/MIST/}} \citep{Choi2016,Dotter2016}. 

The MIST model $G$, $G_\mathrm{RP}$ and $G_\mathrm{BP}$-band magnitudes were converted to colors by finding the extinction in each passband using the following procedure \citep{Babusiaux2018}. First, the reddening $E(B-V)=0.04$ was obtained from the literature \citep{Taylor2007} and was used together with the \citet{Cardelli1989} extinction law to find $A_0=3.1 E(B-V)$. The extinction coefficients for each passband, $k_\lambda = A_\lambda/A_0$, were found using the formula from \citet{Danielski2018} with the numerical coefficients provided by \citet{Babusiaux2018}. The intrinsic color was then corrected for extinction using 
\begin{equation}
(G_\mathrm{BP} - G_\mathrm{RP})_\mathrm{obs} = (G_\mathrm{BP} - G_\mathrm{RP})_0 + A_0(k_\mathrm{BP} - k_\mathrm{RP}) ,
\label{eq:extinct}
\end{equation}
which transforms the MIST model to the M67 color-magnitude diagram. A distance to M\,67 of 830~pc, consistent with the range of reported values 800--900~pc \citep[e.g.,][]{Geller2015} was assumed.  Figure~\ref{cmd-iso-pulse}(a) shows a \textit{Gaia}~Data Release 2 (DR2) photometry color-magnitude diagram (CMD) of M\,67 designated member stars from \citet{Geller2015} \citep[see also][]{Brady2023} together with the MIST isochrone for $\log_{10} \mbox{Age} = 9.6$. 

We transformed our MESA model outputs to the same CMD by converting
the model effective temperature $T_\mathrm{eff}$ to the \textit{Gaia}
color $(G_\mathrm{BP}-G_\mathrm{RP})_0$ using the color-temperature
relation found by \citet{Mucciarelli2020}. Eq.~\ref{eq:extinct} was then used to correct for extinction. The absolute $G$-band
magnitudes were obtained from the model luminosities using the
bolometric-correction polynomial fit of \citet{Andrae2018} and then corrected for distance.

Figure~\ref{cmd-iso-pulse}(b) shows the M\,67 stars together with the segments of the evolutionary tracks for the 
heat-pulse MESA models for the time interval $\log_{10} \mbox{Age} = 9.6\pm 0.0106$. The heat-pulse models reach 
regions of the CMD that are blueward of the Main Sequence and above the subgiant branch. These regions contain 
some M67 designated binary stars. Figure~\ref{cmd-bin-pulse}(a) shows the partial evolutionary tracks for the 
heat-pulse models against the MIST isochrone. The evolutionary tracks agree well with the M\,67 member stars and 
with the MIST isochrone. During the heat pulse, the model stars reach regions of the CMD that are slightly bluer 
and  above the main-sequence and subgiant stretches of the isochrone. These models return immediately to their 
original tracks after the $10^7$~yr pulse. Our highest mass model, $1.32\,M_\odot$, is at the base of the red 
giant branch when the heat pulse occurs. The pulse lifts it blueward to a new trajectory at higher magnitude, 
which then tracks towards the red giant branch. Once the pulse is over it descends the red giant branch to its 
original position. It is possible that mass loss occurs at the highest point but this is beyond the scope of this 
paper. The effect of the $0.8\,M_\odot$ companion on the position of the primary in the CMD is depicted in 
Fig.~\ref{cmd-bin-pulse}(b). The companion is a low-luminosity, red star and it has little effect on the model stars, 
moving them slightly to the red at marginally higher luminosity. The binary systems remain consistent with both the 
isochrone and the observed M67 designated member stars.

\section{Discussion}
\label{sec:disc}
The tidal heating that was derived for the parameter space chosen for this paper offers an
explanation for objects that lie above the sub-giant branch and to the immediate left of the lower red 
giant branch, as well as for BSSs that lie blueward of the Main Sequence but very close to it.  
This would mean that these objects are in a very short-lived evolutionary state ($\sim$0.01~Gyr).  
Extending the parameter space to more massive companions, closer orbital separations and faster or
slower rotation ($\beta_0>$1.3 or $\beta_0<$0.7) would produce higher rates of tidal heating, thus 
potentially extending the pulse-heated tracks to even bluer locations where the majority of BSS stars 
lie. A future investigation will explore this possibility. 

It is also important to note that within the merger scenario for BSS, the progenitors would necessarily 
transit through a phase of non-synchronous rotation and be tidally heated.  They would therefore appear 
as BSS's prior to the merger. 

Finally, it is tempting to speculate that oscillations during the pulse-heated phase, if real, could 
trigger significant mass-loss, stripping the outer stellar layers.  Thus, tidal heating is expected 
to play an important role in all of the BSS formation channels currently under analysis.

Many known binary BSSs are reported to have white dwarf (WD) companions in very long periods. However,
this does not exclude the possibility that they are triple systems in which the BSS forms part of
an inner system whose binary nature is as yet to be revealed.  Other systems do have shorter orbital periods. 
\citet{2023MNRAS.524.1360V} studied the \textit{K2} and \textit{TESS} data of five BSSs in M67 that 
are known to be spectroscopic binaries.  They estimate an orbital period for WOCS~1007
$P_\mathrm{orb}=4.2$~d, $e=0.21$, $m_1=1.95\pm0.26\,M_\odot$ (similar to what is obtained from 
isochrones), $m_2=0.22\pm0.05\,M_\odot$, $R_2=0.07\,R_\odot$8 (confirming the secondary to be a 
low-mass WD, which could not have formed through a single-star evolution scenario). 
\citet{2023MNRAS.524.1360V} postulate that the progenitor system of WOCS~1007 was an almost 
equal-mass ($\sim 1.35\,M_\odot$) binary that has experienced mass transfer. WOCS~1007 is 
known to pulsate but this is due to its location in the $\delta$~Scuti instability region 
rather than tidally induced perturbations.
Although no eclipses were found in the light curves of WOCS~5005 and 1025 (see Fig.~\ref{cmd-iso-pulse} 
for the locations of these systems in the M67 color-magnitude diagram), thereby ruling out 
high-inclination close binary systems, it does not exclude the possibility of low-inclination 
close binary companions \citep{2023MNRAS.524.1360V}. No pulsations were found in these systems. 
The authors report that WOCS~4003 system was more difficult to interpret, with the data suggesting a 
possible low-inclination compact triple system. Detailed light-curve analysis, such as that carried 
out by \citet{2023MNRAS.524.1360V}, is a powerful technique for constraining the properties of 
compact binary systems, which is essential for understanding the formation scenarios of BSS.

\section{Summary and Conclusions} \label{sec:conclusions}

The HRDs of stellar clusters generally display a small number of stars that are located far from 
the cluster isochrone.  
The most striking examples are the Blue Straggler Stars (BSSs) which lie to the blue of the main 
sequence and tend to have higher luminosities than the main sequence turnoff. These anomalies
indicate that the stellar properties have been influenced by physical processes that are not included 
in the standard structure and evolution models. In this paper we tested the hypothesis that stars 
with anomalous  locations on the HRD are the result of energy that is released as heat in their 
differentially rotating shearing layers. We obtained the rate of energy dissipation from an {\it ab initio}
calculation of the tidal perturbations induced by a binary companion and injected it into MESA 
stellar evolution models to study the star's response to the injected heat.  The surface temperature 
and luminosity were transformed to the \textit{Gaia} colors and magnitudes and plotted on the 
color-magnitude diagram of the old cluster M67, which has a $1.26 M_\odot$ MSTO mass. 

We showed that if the dissipated energy is converted into heat, the stellar structure is considerably
altered with respect to that of the standard models.  In particular, the surface temperature and
luminosity are larger, as is the stellar radius.  Thus, the star's location on the HRD is shifted
to higher temperatures and luminosities, similar to what is observed in some of the BSSs.

We studied asynchronously rotating stars in the mass range 1.16-1.32~$M_\odot$ having have a 0.8~$M_\odot$ companion in a 1.44\,d orbit. We found that the heating from the induced differential
rotation has the potential to shift the location of stars in the M67 color-magnitude diagram into several 
anomalous locations, one of which is the blue edge of the main sequence below the turnoff, where
it could broaden the main sequence. A second one is the subgiant branch, where it can increase the
luminosities of stars above the locus of single-stars, and a third one is the lower giant branch where
it  can shift stars to hotter temperatures. Stars with different masses, rotation velocities and
orbital periods could produce objects in yet other anomalous color-magnitude locations. 

Stars that are most likely to undergo heating events are those near the main sequence turnoff,
which is when they start  expanding and changing their internal rotation structure. Thus,
most of the peculiar objects are likely to have masses close to the turnoff mass.  However,
stars with lower masses (still on the MS) or more evolved stars of higher masses could undergo a 
tidal heating event if they are severely perturbed by a tidal capture, leading to the formation of 
a binary system or triple system.  The tidal heating mechanism is universal and is expected to apply 
to stars of all masses.  

Models with the largest heating rates develop large amplitude oscillations
which are tempting to suggest could drive significant mass-loss.  Given the intrinsically asymmetrical
properties of tidal heating, such mass-loss during late phases of stellar evolution could give rise 
to some of the morphologies observed in planetary nebulae.  

Finally, we note that our 1.2~$M_\odot$ pulsed heated model shows a radius increase from 1.50 to 1.75~$R_\odot$ 
on a timescale $<10^3$~yr.  This assumes a constant amount of heat injection but as the radius increases,
so does $\dot{E}$.  Thus, the rise during the tidal heating event could be observed on a significantly shorter 
timescales. Such an event would be interpreted as a relatively slow-rising eruption or outburst,
such as observed in some luminous blue variables on the high-mass end of the HRD or Red Novae
in the low-mass.  Furthermore, because the bloating induced by $\dot{E}$ ultimately comes from
the orbital energy, the orbit will shrink.  As suggested by \citet{2016RMxAA..52..113K}, this could
lead to a runaway process in which the bloating star absorbs a systematically increasing amount
of orbital  energy.  One possible outcome is the ejection of a shell, allowing the star to contract
and stabilize. The alternative would be a merger, in which case the progenitors would transit
through a phase of non-synchronous rotation and be tidally heated.  They would therefore
appear at anomalous HRD locations prior to the merger.

\begin{acknowledgments}
We acknowledge funding for this project from the Institute for Advanced Study of the Indiana 
University Bloomington and UNAM DGAPA/PAPIIT grant 105723. GK expresses gratitude to 
Norbert Langer for numerous enlightening discussions and to the Astronomy Department of Indiana
University for hosting the sabbatical leave during which this research was
conducted.  Software: This research made use of MESA: Modules for Experiments 
in Stellar Astrophysics, r15140//docs.mesastar.org/en/r15140/, \citep{{2011ApJS..192....3P},
{2013ApJS..208....4P}, {2015ApJS..220...15P},{2018ApJS..234...34P},{2019ApJS..243...10P}} and
the rutine mkipp written by Pablo Marchant (https://github.com/orlox/mkipp)
\end{acknowledgments}
\bibliography{TIDES_BSS_2024mar13.bib}{}
\bibliographystyle{aasjournal}

\begin{appendix}

\section{Energy dissipation rates and Kippenhahn diagrams for selected models}

The tidal shear energy dissipation rates in each shell of the 1.2\,M$_\odot$ star for
the h10 model when it has attained a radius of 1.20, 1.32 and 1.50 $R_\odot$ are given
in Table \ref{table_ptk35_a} for a supersynchronously rotating star ($\beta_0$=1.1) with
a 0.4 $M_\odot$ companion, and in Table \ref{table_ptk35_b} for a subsynchronously rotating
star ($\beta_0$=0.7) with a 0.8 $M_\odot$ companion.  The ages at which the radii of the latter 
model are reached, according to the heated MESA model, are listed in Table \ref{ages_radii}.
The radius that is listed in each $r$ column is given in $R_\odot$ and corresponds to the midpoint
of the layer. $\dot{E}$(r) is computed with Eq. (3) and is given in ergs s$^{-1}$. 

Kippenhahn diagrams for the 1.2, 1.26 and 1.32 $M_\odot$ pulse-heated models
The left-hand panels show the stellar interior up to 3.95 Gyr, while the right-hand panels cover the age
range between 3.95 and 4.25 Gyr, which includes the period of energy injection into the outer layers. Note
that the radius scale is different between the two sets of panels. The blue shaded areas are regions of
active nuclear burning, with the shade of blue corresponding to the nuclear energy generation rate, as
indicated in the side bar. The inclined lines indicate the regions where convection occurs, while red
indicates semi-convection zones. The black line represents the stellar surface. 

\begin{deluxetable}{ll|ll|ll|ll}
\tablecaption{Energy dissipation rates for $m_1=1.2 M_\odot$, $\beta_0=1.1$ and $m_2=0.4 M_\odot$ \label{table_ptk35_a}}
\tablecolumns{8}
\tablewidth{0pt}
\tablehead{\multicolumn{2}{c}{$R_1=0.97 R_\odot$} &\multicolumn{2}{c}{$R_1=1.20 R_\odot$}&\multicolumn{2}{c}{$R_1=1.32 R_\odot$}&\multicolumn{2}{c}{$R_1=1.50 R_\odot$}\\
\colhead{$r$}&$\dot{E}(r)$&\colhead{$r$}&$\dot{E}(r)$&\colhead{$r$}&$\dot{E}(r)$&\colhead{$r$}&$\dot{E}(r)$}
\startdata
 0.42 &  1.68 10$^{29}$& 0.52&   3.09 10$^{29}$& 0.57 &  1.69 10$^{29}$& 0.65&   2.65 10$^{29}$ \\
 0.48 &  2.23 10$^{29}$& 0.59&   4.18 10$^{29}$& 0.65 &  3.73 10$^{29}$& 0.74&   4.02 10$^{29}$ \\
 0.53 &  1.50 10$^{29}$& 0.66&   3.01 10$^{29}$& 0.73 &  3.10 10$^{29}$& 0.82&   8.96 10$^{29}$ \\
 0.59 &  2.78 10$^{29}$& 0.73&   8.11 10$^{29}$& 0.81 &  6.72 10$^{29}$& 0.92&   8.69 10$^{29}$ \\
 0.65 &  6.44 10$^{29}$& 0.80&   1.52 10$^{30}$& 0.88 &  2.56 10$^{30}$& 1.00&   4.57 10$^{30}$ \\
 0.71 &  1.84 10$^{30}$& 0.88&   4.89 10$^{30}$& 0.96 &  8.14 10$^{30}$& 1.10&   1.98 10$^{31}$ \\
 0.77 &  2.55 10$^{30}$& 0.95&   1.18 10$^{31}$& 1.04 &  2.67 10$^{31}$& 1.19&   6.56 10$^{31}$ \\
 0.82 &  1.60 10$^{30}$& 1.02&   1.85 10$^{31}$& 1.12 &  4.94 10$^{31}$& 1.28&   1.29 10$^{32}$ \\
 0.88 &  2.72 10$^{30}$& 1.09&   4.48 10$^{31}$& 1.20 &  1.27 10$^{32}$& 1.37&   4.26 10$^{32}$ \\
 0.94 &  6.47 10$^{30}$& 1.16&   1.96 10$^{31}$& 1.28 &  4.79 10$^{31}$& 1.46&   2.34 10$^{32}$ \\
\enddata
\end{deluxetable}

\begin{deluxetable}{ll|ll|ll|ll}
\tablecaption{Energy dissipation rates for $m_1=1.2 M_\odot$, $\beta_0=0.7$ and $m_2=0.8 M_\odot$ \label{table_ptk35_b}}
\tablecolumns{8}
\tablewidth{0pt}
\tablehead{\multicolumn{2}{c}{$R_1=0.97~R_\odot$} &\multicolumn{2}{c}{$R_1=1.20~R_\odot$}&\multicolumn{2}{c}{$R_1=1.32~R_\odot$}&\multicolumn{2}{c}{$R_1=1.50~R_\odot$}\\
\colhead{$r$}&$\dot{E}(r)$&\colhead{$r$}&$\dot{E}(r)$&\colhead{$r$}&$\dot{E}(r)$&\colhead{$r$}&$\dot{E}(r)$}
\startdata
 0.42 &  1.91 10$^{29}$& 0.52&   4.10 10$^{29}$& 0.57 &  2.90 10$^{29}$& 0.65&   7.35 10$^{29}$ \\
 0.48 &  2.90 10$^{29}$& 0.59&   6.08 10$^{29}$& 0.65 &  7.34 10$^{29}$& 0.74&   2.16 10$^{30}$ \\
 0.53 &  4.77 10$^{29}$& 0.66&   2.24 10$^{30}$& 0.73 &  5.48 10$^{30}$& 0.82&   1.90 10$^{31}$ \\
 0.59 &  2.48 10$^{30}$& 0.73&   2.86 10$^{31}$& 0.81 &  6.51 10$^{31}$& 0.92&   1.92 10$^{32}$ \\
 0.65 &  1.56 10$^{31}$& 0.80&   1.88 10$^{32}$& 0.88 &  3.90 10$^{32}$& 1.00&   9.36 10$^{32}$ \\
 0.71 &  5.62 10$^{31}$& 0.88&   6.26 10$^{32}$& 0.96 &  1.18 10$^{33}$& 1.10&   2.52 10$^{33}$ \\
 0.77 &  1.07 10$^{32}$& 0.95&   1.21 10$^{33}$& 1.04 &  2.12 10$^{33}$& 1.19&   4.22 10$^{33}$ \\
 0.82 &  1.44 10$^{32}$& 1.02&   1.59 10$^{33}$& 1.12 &  3.01 10$^{33}$& 1.28&   7.30 10$^{33}$ \\
 0.88 &  4.27 10$^{32}$& 1.09&   4.42 10$^{33}$& 1.20 &  1.09 10$^{34}$& 1.37&   3.93 10$^{34}$ \\
 0.94 &  3.03 10$^{32}$& 1.16&   1.97 10$^{33}$& 1.28 &  6.44 10$^{33}$& 1.46&   2.98 10$^{34}$ \\
\enddata
\end{deluxetable}

\begin{figure}
\includegraphics[width=0.95\columnwidth]{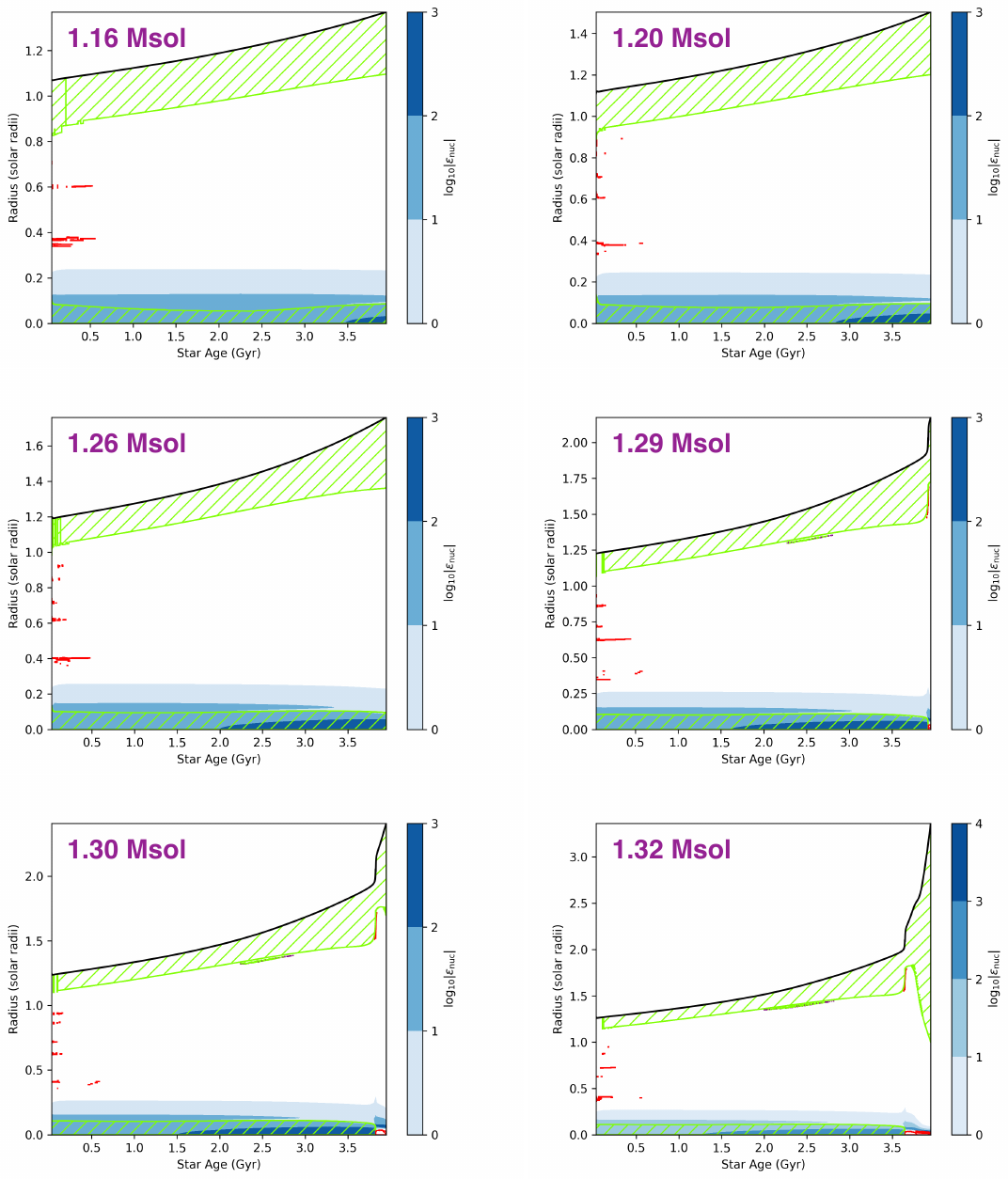}
\caption{Kippenhahn diagrams for the 1.2, 1.26 and 1.32 $M_\odot$ pulse-heated models.
These diagrams were produced with the mkipp software authored by Pablo Marchant https://github.com/orlox/mkipp.
}
\label{kipp_pulsed}
\end{figure}

\end{appendix}

\end{document}